\documentclass[global]{journal}
\usepackage{times}
\usepackage{ccl}

\journalname{Acta Informatica}

\begin{document}

\title{Machine Structure Oriented Control Code Logic}
\subtitle{(Extended Version)}

\author{J.A. Bergstra \and C.A. Middelburg}

\institute{Programming Research Group,
           University of Amsterdam, \\
           Science Park~107, 1098~XG~Amsterdam, the Netherlands\\
           \email{J.A.Bergstra@uva.nl, C.A.Middelburg@uva.nl}}

\maketitle

\begin{abstract}
Control code is a concept that is closely related to a frequently
occurring practitioner's view on what is a program: code that is capable
of controlling the behaviour of some machine.
We present a logical approach to explain issues concerning control codes
that are independent of the details of the behaviours that are
controlled.
Using this approach, such issues can be explained at a very abstract
level.
We illustrate this among other things by means of an example about the
production of a new compiler from an existing one.
The approach is based on abstract machine models, called machine
structures.
We introduce a model of systems that provide execution environments for
the executable codes of machine structures and use it to go into
portability of control codes.
\end{abstract}

\section{Introduction}
\label{sect-introduction}

In theoretical computer science, the meaning of programs usually plays a
prominent part in the explanation of many issues concerning programs.
Moreover, what is taken for the meaning of programs is mathematical by
nature.
On the other hand, it is customary that practitioners do not fall back
on the mathematical meaning of programs in case explanation of issues
concerning programs is needed.
More often than not, they phrase their explanations from the viewpoint
that a program is code that is capable of controlling the behaviour of
some machine.
Both theorists and practitioners tend to ignore the existence of this
contrast.
In order to break through this, we as theorists make in this paper an
attempt to map out the way in which practitioners explain issues
concerning programs.

We informally define control code as code that is capable of controlling
the behaviour of some machine.
There are control codes that fail to qualify as programs according to
any conceivable theory of programming.
For that reason, we make the distinction between control codes and
programs.
However, there are issues concerning programs that can be explained at
the level of control codes by considering them as control codes that
qualify as programs.
Relative to a fixed machine, the machine-dependent concept of control
code that qualifies as program is more abstract than the
machine-independent concept of program: control code that qualifies as
program is just representative (on the fixed machine) of behaviour
associated with a program with which it is possible to explain the
behaviour.
This might be an important motive to explain issues concerning programs
at the level of control codes.

To simplify matters, we consider in this paper non-interactive behaviour
only.
We consider this simplification desirable to start with.
Henceforth, control codes are implicitly assumed to control
non-interactive behaviour only and the behaviours \sloppy associated
with programs are implicitly assumed to be non-interactive.

Our attempt to map out the way in which practitioners explain issues
concerning programs yields a logical approach to explain issues
concerning control codes that are independent of the details of the
behaviours that are controlled.
Machine structures are used as a basis of the approach.
They are inspired by the machine functions introduced in~\cite{ES70a}
to provide a mathematical basis for the T-diagrams proposed
in~\cite{Bra61a}.
A machine structure offers a machine model at a very abstract level.

We illustrate the approach by means of some examples.
The issues explained in the examples are well understood for quite a
time.
They are primarily meant to demonstrate the effectiveness of the
approach.
In the explanations given, we have consciously been guided by empirical
viewpoints usually taken by practitioners rather than theoretical
viewpoints.
Those empirical viewpoints may be outside the perspective of some
theorists.

Mapping out the way in which practitioners explain issues concerning
programs, phrased as a matter of applied mathematics, seems to lead
unavoidably to unexpected concepts and definitions.
This means among other things that steps made in this paper cannot
always be motivated directly from the practice that we map out.
This is an instance of a general property of applied mathematics that we
have to face: the design of a mathematical theory does not follow
imperatively from the problems of the application area concerned.

We believe that the presented approach is useful because in various
areas frequently no distinction is made between programs and control
codes and interest is primarily in issues concerning control codes that
are independent of the details of the behaviours that are controlled.
Some examples of such areas are software asset sourcing and software
patents.%
\footnote
{Software asset sourcing is an important part of IT sourcing,
 see e.g.~\cite{LV92a,Ver05a,Del07a}.
 An extensive study of software patents and their implications on
 software engineering practices can be found in~\cite{BK06a}.
}
Moreover, we find that control code production is in the end what
software construction is about.

Machine structures in themselves are not always sufficient to explain
issues concerning control codes that are independent of the details of
the behaviours that are controlled.
If systems that provide execution environments for the executable codes
of machine structures are involved, then more is needed.
We introduce an execution architecture for machine structures as a model
of such systems and explain portability of control codes using this
execution architecture.
An extension of basic thread algebra, introduced in~\cite{BL02a} under
the name basic polarized process algebra, is used to describe processes
that operate upon the execution architecture.
The reason to use basic thread algebra is that it has been designed as
an algebra of processes that interact with machines of the kind to which
also the execution architecture belongs.
It is quite awkward to describe processes of that kind using a general
process algebra such as ACP~\cite{Fok00}, CCS~\cite{Mil89} or
CSP~\cite{Hoa85}.

This paper is organized as follows.
First, we introduce machine structures
(Section~\ref{sect-machine-structures}).
Next, we introduce control code notations and program notations
(Section~\ref{sect-cc-and-prg}).
Then, we present our approach to explain issues concerning control codes
by means of examples about the production of a new assembler using an
existing one and the production of a new compiler using an existing one
(Section~\ref{sect-assembler-compiler}).
We also use this approach to explain the relation between compilers and
interpreters (Section~\ref{sect-interpreter}).
Following this, we sum up the effects of withdrawing a simplifying
assumption concerning the representation of control codes made in the
foregoing (Section~\ref{sect-bs-repr-cc}).
After that, we outline an execution architecture for machine structures
(Section~\ref{sect-exearch}).
Then, we review the extension of basic thread algebra that covers the
effects of applying threads to services (Section~\ref{sect-TA}).
Following this, we formalize the execution architecture for machine
structures and define the family of services determined by it
(Section~\ref{sect-exearch-formal}).
After that, we explain portability of control codes using thread algebra
and the execution architecture services (Section~\ref{sect-cc-exearch}).
Finally, we make some concluding remarks (Section~\ref{sect-concl}).

Up to Section~\ref{sect-exearch}, this paper is a major revision
of~\cite{Ber04a}.
It has been substantially rewritten so as to streamline the material.
Several important technical aspects have been significantly modified.

\section{Machine Functions and Machine Structures}
\label{sect-machine-structures}

In this section, machine structures are introduced.
Machine structures are the basis for our approach to explain issues
concerning control codes.
They are very abstract machine models and cover non-interactive machine
behaviour only.

First, we introduce the notion of machine function introduced
in~\cite{Ber04a}.
It generalizes the notion of machine function introduced in~\cite{ES70a}
by covering machines with several outputs.
Machine functions are very abstract machine models as well, but they are
less suited than machine structures to model general purpose machines
such as computers.
Machine structures can easily be defined without reference to machine
functions.
The introduction of machine functions is mainly for expository reasons.

\subsection{Machine Functions}
\label{subsect-machine-functions}

A machine function $\mf$ is actually a family of functions: it consists
of a function $\mf_n$ for each natural number $n > 0$.
Those functions map each finite sequence of bit sequences to either a
bit sequence or $\Mea$ or $\Div$.
Here, $\Mea$ stands for meaningless and $\Div$ stands for divergent.
A machine function is supposed to model a machine that takes several bit
sequences as its inputs and produces several bit sequences as its
outputs unless it does not halt on the inputs.
Let $x_1,\ldots,x_m$ be bit sequences.
Then the connection between the machine function $\mf$ and the machine
modelled by it can be understood as follows:%
\footnote
{We write
 $\emptyseq$ for the empty sequence,
 $\seq{x}$ for the sequence having $x$ as sole element, and
 $\chi \conc \chi'$ for the concatenation of finite sequences $\chi$
 and $\chi'$.
 We use $\seq{x_1,\ldots,x_n}$ as a shorthand for
 $\seq{x_1} \conc \ldots \conc \seq{x_n}$.
 We write $\seqof{X}$ for the set of all finite sequences with elements
 from set $X$.
}
\begin{iteml}
\item
if $\mf_n(\seq{x_1,\ldots,x_m})$ is a bit sequence, then the machine
function $\mf$ models a machine that produces
$\mf_n(\seq{x_1,\ldots,x_m})$ as its $n$th output on it taking
$x_1$, \ldots, $x_m$ as its inputs;
\item
if $\mf_n(\seq{x_1,\ldots,x_m})$ is $\Mea$, then the machine function
$\mf$ models a machine that produces less than $n$ outputs on it taking
$x_1$, \ldots, $x_m$ as its inputs;
\item
if $\mf_n(\seq{x_1,\ldots,x_m})$ is $\Div$, then the machine function
$\mf$ models a machine that does not produce any output on it taking
$x_1$, \ldots, $x_m$ as its inputs because it does not halt on the
inputs.
\end{iteml}
Concerning the machine modelled by a machine function, we assume the
following:
\begin{iteml}
\item
if it does not halt, then no output gets produced;
\item
if it does halt, then only finitely many outputs are produced;
\item
if it does not halt, then this cannot be prevented by providing more
inputs;
\item
if it does halt, then the number of outputs cannot be increased by
providing less inputs.
\end{iteml}
The intuitions behind the first two assumptions are obvious.
The intuition behind the third assumption is that, with respect to not
halting, a machine does not use more inputs than it needs.
The intuition behind the last assumption is that, with respect to
producing outputs, a machine does not use more inputs than it needs.

Henceforth, we write $\BitSeq$ for the set $\seqof{\set{0,1}}$ of
\emph{bit sequences}.
It is assumed that $\Mea \not\in \BitSeq$ and $\Div \not\in \BitSeq$.

We now define machine functions in a mathematically precise way.

Let $\Bseq \subseteq \BitSeq$.
Then a \emph{machine function} $\mf$ on $\Bseq$ is a family of functions
\begin{ldispl}
\indfam
 {\funct{\mf_n}{\seqof{\Bseq}}{(\Bseq \union \set{\Div,\Mea})}}
 {n \in \Nat}
\end{ldispl}%
satisfying the following rules:
\begin{ldispl}
\AND{n \in \Nat}
 \bigl(\AND{m \in \Nat}
        (\mf_n(\chi) = \Div \Implies \mf_m(\chi) = \Div)\bigr)\;,
\\
\AND{n \in \Nat}
 \bigl(\mf_n(\chi) \neq \Div \Implies
       \bigl(\OR{m \in \Nat, m > n}
               \mf_m(\chi) = \Mea\bigr)\bigr)\;,
\\
\AND{n \in \Nat}
 \bigl(\AND{m \in \Nat, m > n}
        (\mf_n(\chi) = \Mea \Implies \mf_m(\chi) = \Mea)\bigr)\;,
\\
\AND{n \in \Nat}
 (\mf_n(\chi) = \Div \Implies \mf_n(\chi \conc \chi') = \Div)\;,
\\
\AND{n \in \Nat}
 (\mf_n(\chi \conc \chi') = \Mea \Implies \mf_n(\chi) = \Mea)\;.
\end{ldispl}%
We write $\MF$ for the set of all machine functions.

\begin{example}
\label{example-machine-function}
Take a high-level programming language $\nm{PL}$ and an assembly
language $\nm{AL}$.
Consider a machine function $\nm{cf}$, which models a machine dedicated
to compiling $\nm{PL}$ programs, and a machine function $\nm{df}$, which
models a machine dedicated to disassembling executable codes.
Suppose that the compiling machine takes a bit sequence representing
a $\nm{PL}$ program as its only input and produces a bit sequence
representing an $\nm{AL}$ version of the $\nm{PL}$ program as its first
output, a bit sequence representing a listing of error messages as its
second output, and an executable code for the $\nm{PL}$ program as its
third output.
Moreover, suppose that the disassembling machine takes an executable
code as its only input and producing a bit sequence representing an
$\nm{AL}$ version of the executable code as its first output and a
bit sequence representing a listing of error messages as its second
output.
The relevant properties of the machines modelled by $\nm{cf}$ and
$\nm{df}$ that may now be formulated include:
\begin{ldispl}
\nm{cf}_2(\seq{x}) = \emptyseq \Implies
\nm{cf}_1(\seq{x}) \neq \emptyseq\;,
\\
\nm{df}_2(\seq{x}) = \emptyseq \Implies
\nm{df}_1(\seq{x}) \neq \emptyseq\;,
\\
\nm{cf}_2(\seq{x}) = \emptyseq \Implies
\nm{df}_1(\nm{cf}_3(\seq{x})) = \nm{cf}_1(\seq{x})\;.
\end{ldispl}%
These formulas express that executable code is produced by the compiling
machine unless errors are found, disassembly succeeds unless errors are
found, and disassembly is the inverse of assembly.
\end{example}

Machines such as the compiling machine and the disassembling machine are
special purpose machines.
They are restricted to exhibit a particular type of behaviour.
Computers are general purpose machines that can exhibit different types
of behaviour at different times.
This is possible because computers are code controlled machines.
A code controlled machine takes one special input that controls its
behaviour.
In general, not all bit sequences that a code controlled machine can
take as its inputs are capable of controlling the behaviour of that
machine.
The bit sequences that are capable of controlling its behaviour are
known as its executable codes.
Note that executable code is a machine-dependent concept.

Machine functions can be used to model code controlled machines as well.
We will use the phrase code controlled machine function for machine
functions that are used to model a code controlled machine.
We will use the convention that the first bit sequence in the argument
of the functions that make up a code controlled machine function
corresponds to the special input that controls the behaviour of the
machine modelled.
Because, in general, not all bit sequences that a code controlled
machine can take as its inputs are executable codes, more than just a
machine function is needed to model a code controlled machine.
That is why we introduce machine structures.

\subsection{Machine Structures}
\label{subsect-machine-structures}

A machine structure $\gM$ consists a set of bit sequences $\Bseq$,
functions $\mf_n$ that make up a machine function on $\Bseq$, and a
subset $\Exec$ of $\Bseq$.
If $\Exec$ is empty, then the machine structure $\gM$ is essentially the
same as the machine function contained in it.
If $\Exec$ is not empty, then the machine structure $\gM$ is supposed to
model a code controlled machine.
In the case where $\Exec$ is not empty, the connection between the
machine structure $\gM$ and the code controlled machine modelled by it
can be understood as follows:
\begin{iteml}
\item
$\Bseq$ is the set of all bit sequences that the code controlled machine
modelled by $\gM$ can take as its inputs;
\item
if $x \in \Exec$, then the bit sequence $x$ belongs to the executable
codes of the code controlled machine modelled by $\gM$;
\item
if $x \in \Exec$, then the functions $\mf'_n$ that are defined by
$\mf'_n(\seq{y_1,\ldots,y_m}) = \mf_n(\seq{x,y_1,\ldots,y_m})$ make up
a machine function on $\Bseq$ modeling a machine that exhibits the same
behaviour as the code controlled machine modelled by $\gM$ exhibits
under control of the executable code $x$.
\end{iteml}
The assumptions made about the machine modelled by a machine structure
are the same as the assumptions made before about the machine modelled
by a machine function.
It is tempting to add the following assumption:
\begin{iteml}
\item
if the special input meant to control its behaviour does not belong to
its executable codes, then the machine halts without having produced any
output.
\end{iteml}
We refrain from adding this assumption because it is to be expected
that:
(a)~we can do without it in explaining issues concerning control codes;
(b)~it does not hold good for all machines that we may encounter.
Moreover, in case we would incorporate this assumption in the notion of
machine structure, it would not supersede the notion of machine
function.

We now define machine structures in a mathematically precise way.

A \emph{machine structure} $\gM$ is a structure composed of
\begin{iteml}
\item
a set $\Bseq \subseteq \BitSeq$,
\item
a unary function
$\funct{\mf_n}{\seqof{\Bseq}}{(\Bseq \union \set{\Div,\Mea})}$,
for each $n \in \Nat$,
\item
a unary relation $\Exec \subseteq \Bseq$,
\end{iteml}
where the family of functions
$\indfam
  {\funct{\mf_n}{\seqof{\Bseq}}{(\Bseq \union \set{\Div,\Mea})}}
  {n \in \Nat}$
is a machine function on $\Bseq$.
We say that $\gM$ is a \emph{code controlled machine structure} if
$\Exec \neq \emptyset$, and we say that $\gM$ is a
\emph{dedicated machine structure} if $\Exec = \emptyset$.

Let $\gM = \tup{\Bseq,\indfam{\mf_n}{n \in \Nat},\Exec}$ be a code
controlled machine structure, and let $x \in \Exec$.
Then the \emph{meaning} of $x$ with respect to $\gM$, written
$\bextr{x}{\gM}$, is the machine function
\begin{ldispl}
\indfam
  {\funct{\mf'_n}{\seqof{\Bseq}}{(\Bseq \union \set{\Div,\Mea})}}
  {n \in \Nat}\;,
\end{ldispl}%
where the functions $\mf'_n$ are defined by
\begin{ldispl}
\mf'_n(\seq{y_1,\ldots,y_m}) = \mf_n(\seq{x,y_1,\ldots,y_m})\;.
\end{ldispl}%
Moreover, let $x',x'' \in \Exec$.
Then $x'$ is \emph{behaviourally equivalent} to $x''$ on $\gM$, written
$x' \beqv{\gM} x''$, if $\bextr{x'}{\gM} = \bextr{x''}{\gM}$.

Let $\gM = \tup{\Bseq,\indfam{\mf_n}{n \in \Nat},\Exec}$ be a code
controlled machine structure.
Then we will write
\begin{ldispl}
\ccmf{\gM}{n}{x}{y_1,\ldots,y_m}\quad \mathrm{for}\quad
\mf_n(\seq{x,y_1,\ldots,y_m})\;.
\end{ldispl}%
Moreover, we will write
\begin{ldispl}
\ccmf{\gM}{{}}{x}{y_1,\ldots,y_m}\quad \mathrm{for}\quad
\ccmf{\gM}{1}{x}{y_1,\ldots,y_m}\;.
\end{ldispl}%
We will also omit $\gM$ if the machine structure is clear from the
context.

\begin{example}
\label{example-machine-structure}
Take a code controlled machine structure
$\gM = \tup{\Bseq,\indfam{\mf_n}{n \in \Nat},\Exec}$.
Consider again the machine functions $\nm{cf}$ and $\nm{df}$ from
Example~\ref{example-machine-function}.
These machine functions model a machine dedicated to compiling programs
in some high-level programming language $\nm{PL}$ and a machine
dedicated to disassembling executable codes, respectively.
Let $e_\nm{cf},e_\nm{df} \in E$ be such that
\begin{ldispl}
\bextr{e_\nm{cf}}{\gM} = \nm{cf}\quad \mathrm{and}\quad
\bextr{e_\nm{df}}{\gM} = \nm{df}\;.
\end{ldispl}%
Then $e_\nm{cf}$ and $e_\nm{df}$ are executable codes that control the
behaviour of the code controlled machine modelled by $\gM$ such that
this machine behaves the same as the dedicated machine modelled by
$\nm{cf}$ and the dedicated machine modelled by $\nm{df}$, respectively.
This implies that for all $x \in \Bseq$ and $n \in \Nat$:
\begin{ldispl}
\ccmf{\gM}{n}{e_\nm{cf}}{x} = \nm{cf}_n(\seq{x})\quad \mathrm{and}\quad
\ccmf{\gM}{n}{e_\nm{df}}{x} = \nm{df}_n(\seq{x})\;.
\end{ldispl}%
Note that for $\nm{cf}$ there may be an $e'_\nm{cf} \in E$ with
$e'_\nm{cf} \neq e_\nm{cf}$ such that
$\bextr{e'_\nm{cf}}{\gM} = \nm{cf}$, and likewise for $\nm{df}$.
\end{example}

A code controlled machine structure
$\gM = \tup{\Bseq,\indfam{\mf_n}{n \in \Nat},\Exec}$ determines all by
itself a machine model.
For an execution, which takes a single step, an executable code
$x \in \Exec$, a sequence $\seq{y_1,\ldots,y_m} \in \seqof{\Bseq}$ of
inputs and the machine function $\indfam{\mf_n}{n \in \Nat}$ are needed.
The executable code is not integrated in the machine in any way.
In particular, it is not stored in the machine.
As nothing is known about any storage mechanism involved, due to the
abstract nature of machine structures, it is not plausible to classify
the model as a stored code machine model.

\subsection{Identifying the Input that Controls Machine Behaviour}
\label{subsect-identification}

It is a matter of convention that the first bit sequence in the argument
of the functions that make up the machine function of a code controlled
machine structure corresponds to the special input that controls the
behaviour of the machine modelled.
The issue is whether a justification for this correspondence can be
found in properties of the code controlled machine structure.
This amounts to identifying the input that controls the behaviour of the
machine modelled.

Take the simple case where always two inputs are needed to produce any
output and always one output is produced.
Then a justification for the correspondence mentioned above can be found
only if the machine function involved is asymmetric and moreover the
first bit sequence in the argument of the function that yields the first
output overrules the second bit sequence.
Here, by overruling is meant being more in control.

In this simple case, the criteria of asymmetry and overruling can easily
be made more precise.
Suppose that $\gM = \tup{\Bseq,\indfam{\mf_n}{n \in \Nat},\Exec}$ is a
code controlled machine structure that models a machine that needs
always two inputs to produce any output and produces always one output.
Then the machine function $\indfam{\mf_n}{n \in \Nat}$ is asymmetric if
there exist $x,y \in \Bseq$ such that $\mf_1(x,y) \neq \mf_1(y,x)$.
The first bit sequence in the argument of the function $\mf_1$ overrules
the second one if there exist $x_1,x_2 \in \Exec$ and
$z_1,z_2 \in \Bseq$ with $z_1 \neq z_2$ such that $\mf_1(x_1,y) = z_1$
and $\mf_1(x_2,y) = z_2$ for all $y \in \Bseq$.
It is easily proved that the first bit sequence in the argument of the
function $\mf_1$ overrules the second one only if the second bit
sequence in the argument of the function $\mf_1$ does not overrule the
first one.

The criterion of overruling becomes more interesting if more than two
inputs may be needed to produce any output, because this is usually the
case with general-purpose machines.
For example, on a general-purpose machine, the first input may be an
executable code for an interpreter of intermediate codes produced by a
compiler for some high-level programming language $\nm{PL}$, the second
input may be a bit sequence representing the intermediate code for a
$\nm{PL}$ program, and one or more subsequent inputs may be bit
sequences representing data needed by that program.
In this example, the first input overrules the second input and
subsequent inputs present and in addition the second input overrules
the third input and subsequent inputs present.

\section{Control Code Notations and Program Notations}
\label{sect-cc-and-prg}

In this section, we introduce the concepts of control code notation and
program notation in the setting of machine structures and discuss the
differences between these two concepts.
The underlying idea is that a control code is a code that is capable of
controlling the behaviour of some machine and a program is a control
code that is acquired by programming.
The point is that there exist control codes that are not acquired by
programming.
In~\cite{Jan08a}, which appeared after the report version of the current
paper~\cite{BM07bP}, a conceptual distinction is made between proper
programs and dark programs.
We found that proper programs correspond with control codes that are
acquired by programming and dark programs correspond with  control codes
that are not acquired by programming.
As a matter of fact, the notion of machine structure allows for the
discussion of proper and dark programs in~\cite{Jan08a} to be made more
precise.

\subsection{Control Code Notations}
\label{subsect-cc}

The intuition is that, for a fixed code controlled machine, control
codes are objects (usually texts) representing executable codes of that
code controlled machine.
The principal examples of control codes are the executable codes
themselves.
Note that, like the concept of executable code, the concept of control
code is machine-dependent.
A control code notation for a fixed code controlled machine is a
collection of objects together with a function which maps each of the
objects from that collection to a particular executable code of the code
controlled machine.

In order to make a code controlled machine transform members of one
control code notation into members of another control code notation,
like in compiling and assembling, control codes that are not bit
sequences must be represented by bit sequences.
To simplify matters, we will assume that all control code notations are
collections of bit sequences.
Assuming this amounts to identifying control codes with the bit
sequences representing them.
In Section~\ref{sect-bs-repr-cc}, we will withdraw this assumption.

Let $\gM = \tup{\Bseq,\indfam{\mf_n}{n \in \Nat},\Exec}$ be a code
controlled machine structure.
Then a \emph{control code notation} for $\gM$ consists of a set
$\CCN \subseteq \Bseq$ and a function $\funct{\psi}{\CCN}{\Exec}$.
The members of $\CCN$ are called \emph{control codes} for $\gM$.
The function $\psi$ is called a \emph{machine structure projection}.

Let $\tup{\CCN,\psi}$ be a control code notation for a code controlled
machine structure $\tup{\Bseq,\indfam{\mf_n}{n \in \Nat},\Exec}$.
Then we assume that $\psi(c) = c$ for all $c \in \CCN \inter \Exec$.

Let $\gM$ be a code controlled machine structure, let $\tup{\CCN,\psi}$
be a control code notation for $\gM$, and let $c \in \CCN$.
Then the \emph{meaning} of $c$ with respect to $\gM$, written
$\bextrcc{c}{\gM}{\CCN}$, is $\bextr{\psi(c)}{\gM}$.

Control codes, like executable codes, are given a meaning related to one
code controlled machine structure.
The executable codes of a code controlled machine structure themselves
make up a control code notation for that machine structure.
Let $\gM = \tup{\Bseq,\indfam{\mf_n}{n \in \Nat},\Exec}$ be a code
controlled machine structure, and let $1_E$ be the identity function on
$E$.
Then $\tup{E,1_E}$ is a control code notation for $\gM$.
We trivially have $\bextrcc{e}{\gM}{E} = \bextr{e}{\gM}$ for all
$e \in E$.
Henceforth, we loosely write $\Exec$ for the control code notation
$\tup{\Exec,1_\Exec}$.

\subsection{Program Notations}
\label{subsect-programs}

To investigate the conditions under which it is appropriate to say that
a control code notation qualifies as a program notation, it is in fact
immaterial how the concept of program is defined.
However, it is at least convenient to make the assumption that, whatever
the program notation, there is a hypothetical machine model by means of
which the intended behaviour of programs from the program notation can
be explained at a level that is suited to our purpose.
We believe that this assumption is realistic.

Let some theory of programming be given that offers a reliable
definition of the concept of program.
Then an \emph{acknowledged program notation} is a set $\PGN$ of
programs.
It is assumed that there is a well-understood hypothetical machine model
by means of which the intended behaviour of programs from $\PGN$ can be
explained at a level that allows for the input-output relation of
programs from $\PGN$, i.e.\ the kind of behaviour modelled by machine
functions, to be derived.
It is also assumed that this hypothetical machine model determines a
function $\funct{\bextrpg{\ph}{\PGN}}{\PGN}{\MF}$ which maps
programs to the machine functions modelling their behaviour at the
abstraction level of input-output relations.

In~\cite{BL02a}, a theory, called program algebra, is introduced in
which a program is a finite or infinite sequence of instructions.
Moreover, the intended behaviour of instruction sequences is explained
at the level of input-output relations by means of a hypothetical
machine model which involves processing of one instruction at a time,
where some machine changes its state and produces a reply in case the
instruction is not a jump instruction.
This hypothetical machine model is an analytic execution architecture in
the sense of~\cite{BP04a}.
In the current paper, the definition of the concept of program
from~\cite{BL02a} could be used.
However, we have not fixed a particular concept of program because we
intend to abstract from the details involved in any such conceptual
definition.

Note that programs, unlike control codes, are given a meaning using a
hypothetical machine model.
This means that the given meaning is not related to some code controlled
machine structure.

\subsection{Control Code Notations Qualifying as Program Notations}
\label{subsect-qualifying}

The intuition is that a control code notation for a code controlled
machine qualifies as a program notation if there exist an acknowledged
program notation and a function from the control code notation to the
program notation that maps each control code to a program such that, at
the level of input-output relations, the machine behaviour under control
of the control code coincides with the behaviour that is associated with
the corresponding program.
If a control code notation qualifies as a program notation, then its
elements are considered programs.

Let $\gM$ be a code controlled machine structure, and
let $\tup{\CCN,\psi}$ be a control code notation for $\gM$.
Then $\tup{\CCN,\psi}$ \emph{qualifies} as a program notation if there
exist an acknowledged program notation $\PGN$ and a function
$\funct{\phi}{\CCN}{\PGN}$ such that for all $c \in \CCN$:
\begin{ldispl}
\bextr{\psi(c)}{\gM} = \bextrpg{\phi(c)}{\PGN}\;.
\end{ldispl}%

This definition implies that, in the case of a control code notation
that qualifies as a program notation, control codes can be given a
meaning using a hypothetical machine model.
Control code by itself is just representative of machine behaviour
without any indication that it originates from a program with which it
is possible to explain the behaviour by means of a well-understood
hypothetical machine model.
The function $\phi$ whose existence is demanded in the definition is
suggestive of reverse engineering: by its existence, control codes look
to be implementations of programs on a code controlled machine.
We might say that the reason for classifying a control code notation in
the ones that qualify as a program notation lies in the possibility of
reverse engineering.
The function $\phi$ is the opposite of a representation.
It might be called a co-representation.

Suppose that $\gM = \tup{\Bseq,\indfam{\mf_n}{n \in \Nat},\Exec}$ is a
code controlled machine structure and $\tup{E,1_E}$ qualifies as a
program notation.
Then $\gM$ models a code controlled machine whose executable codes
constitute a control code notation that qualifies as a program notation.
Therefore, it is appropriate to call $\gM$ a program controlled machine
structure.
A program controlled machine structure is a code controlled machine
structure, but there is additional information which is considered to
make it more easily understood from the tradition of computer
programming: each executable code can be taken for a program and the
intended behaviour of that program can be explained by means of a
well-understood hypothetical machine model.
It is plausible that, for any code controlled machine structure modeling
a real machine, there is additional information which is considered to
make it more easily understood from some tradition or another.

We take the view that a code controlled machine structure having both
executable codes that can be considered programs and executable codes
that cannot be considered programs are improper.
Therefore, we introduce the notion of proper code controlled machine
structure.

Let $\gM = \tup{\Bseq,\indfam{\mf_n}{n \in \Nat},\Exec}$ be a
code controlled machine structure.
Then $\gM$ is a \emph{proper} code controlled machine structure if
$\tup{E',1_{E'}}$ qualifies as a program notation for some
$E' \subseteq E$ only if $\tup{E,1_E}$ qualifies as a program notation.

\subsection{Control Code Notations Not Qualifying as Program Notations}
\label{subsect-not-qualifying}

The question arises whether all control code notations qualify as
program notations.
If that were true, then the conceptual distinction between control code
notations and program notations is small.
If a control code notation qualifies as a program notation, then all
control codes concerned can be considered the result of implementing a
program on a code controlled machine.
This indicates that counterexamples to the hypothesis that all control
code notations qualify as program notations will concern control codes
that do not originate from programming.
We give two counterexamples where control codes arise from artificial
intelligence.

Consider a neural network in hardware form, which is able to learn while
working on a problem and thereby defining parameter values for many
firing thresholds for artificial neurons.
The parameter values for a particular problem may serve as input for a
machine that needs to address that problem.
These problem dependent parameter inputs can be considered control codes
by all means.
However, there is no conceivable theory of programming according to
which these problem dependent parameter inputs can be considered
programs.
The feature of neural networks that is important here is their ability
to acquire control code by another process than programming.

Consider a purely hardware made robot that processes geographical data
loaded into it to find a target location.
The loaded geographical data constitute the only software that
determines the behaviour of the robot.
Therefore, the loaded geographical data constitute control code.
However, there is no conceivable theory of programming according to
which such control codes can be considered programs.
They are certainly acquired by another process than programming.

In the case of control code notations that qualify as program notations,
the control codes are usually produced by programming followed by
compiling or assembling.
The examples illustrate different forms of control code production
that involve neither programming nor compiling or assembling.
The first example shows that control codes can be produced without
programming by means of artificial intelligence based techniques.
The second example shows that the behaviour of machines applying
artificial intelligence based techniques can be controlled by control
codes that are produced without programming.

\section{Assemblers and Compilers}
\label{sect-assembler-compiler}

In the production of control code, practitioners often distinguish two
kinds of control codes in addition to executable codes: assembly codes
and source codes.
An assembler is a control code corresponding to an executable code of a
code controlled machine that controls the behaviour of that code
controlled machine such that it transforms assembly codes into
executable codes and a compiler is a control code corresponding to an
executable code of a code controlled machine that controls the behaviour
of that code controlled machine such that it that transforms source
codes into assembly codes or executable codes.

In this section, we consider the issue of producing a new assembler
for some assembly code notation using an existing one and the similar
issue of producing a new compiler for some source code notation using an
existing one.
Whether an assembly code notation or a source code notation qualifies as
a program notation is not relevant to these issues.

\subsection{Assembly Code Notations and Source Code Notations}
\label{subsect-assembly-source}

At the level of control codes for machine structures, the control code
notations that are to be considered assembly code notations and the
control code notations that are to be considered source code notations
cannot be characterized.
The level is too abstract.
It happens to be sufficient for many issues concerning assemblers and
compilers, including the ones considered in this section, to simply
assume that some collection of control code notations comprises the
assembly code notations and some other collection of control code
notations comprises the source code notations.

Henceforth, we assume that, for each machine structure $\gM$, disjoint
sets $\AC{\gM}$ and $\SC{\gM}$ of control code notations for $\gM$ have
been given.
The members of $\AC{\gM}$ and $\SC{\gM}$ are called
\emph{assembly code notations} for $\gM$ and
\emph{source code notations} for $\gM$, respectively.

The following gives an idea of the grounds on which control code
notations are classified as assembly code notation or source code
notation.
Assembly code is control code that is very close to executable code.
This means that there is a direct translation of assembly codes into
executable codes.
An assembly code notation is specific to a machine.
Source code is control code that is not very close to executable code.
The translation of source code into executable code is more involved
than the translation of assembly code into executable code.
Usually, a source code notation is not specific to a machine.

A high-level programming language, such as Java~\cite{GJSB00a} or
C\#~\cite{HWG03a}, is considered a source code notation.
The term high-level programming language suggests that it concerns a
notation that qualifies as a program notation.
However, as mentioned above,  whether a source code notation qualifies
as a program notation is not relevant to the issues considered in this
section.

\subsection{Control Code Notations Involved in Assemblers and Compilers}
\label{subsect-assembler-compiler}

Three control code notations are involved in an assembler or compiler:
it lets a code controlled machine transform members of one control code
notation into members of another control code notation and it is itself
a member of some control code notation.
We introduce a special notation to describe this aspect of assemblers
and compilers succinctly.

Let $\gM = \tup{\Bseq,\indfam{\mf_n}{n \in \Nat},\Exec}$ be a code
controlled machine structure, and let $\tup{\CCN,\psi}$,
$\tup{\CCN',\psi'}$ and $\tup{\CCN'',\psi''}$ be control code
notations for $\gM$.
Then we write $\transl{\nm{cc}}{\CCN'}{\CCN''}{\CCN}$
for
\begin{ldispl}
\nm{cc} \in \CCN \And
\Forall{\nm{cc}' \in \CCN'}
 {(\Exists{\nm{cc}'' \in \CCN''}
  {\ccmf{\gM}{{}}{\psi(\nm{cc})}{\nm{cc}'} = \nm{cc}''})}\;.
\end{ldispl}%
We say that $\nm{cc}$ is \emph{in executable form} if
$\CCN \subseteq \Exec$,
that $\nm{cc}$ is \emph{in assembly form} if $\CCN \in \AC{\gM}$, and
that $\nm{cc}$ is \emph{in source form} if $\CCN \in \SC{\gM}$.

\subsection{The Assembler Fixed Point}
\label{subsect-assembler-fix}

In this subsection, we consider the issue of producing a new assembler
for some assembly code notation using an existing one.

Let $\gM = \tup{\Bseq,\indfam{\mf_n}{n \in \Nat},\Exec}$ be a code
controlled machine structure, and let $\tup{\ACN,\psi}$ be a control
code notation for $\gM$ that belongs to $\AC{\gM}$.
Suppose that $\transl{\assemb}{\ACN}{\Exec}{\Exec}$ is an existing
assembler for $\ACN$.
This assembler is in executable form.
Suppose further that a new assembler
$\transl{\assemb'}{\ACN}{\Exec}{\ACN}$ for $\ACN$ is made
available.
This new assembler is not in executable form.
It needs to be assembled by means of the existing assembler.
The new assembler is considered correct if behaviourally equivalent
executable codes are produced by the existing assembler and the one
obtained by assembling the new assembler by means of the existing
assembler,
i.e.\
\begin{equation}
\label{eqn-correct-assembler}
\Forall{\ac \in \ACN}
{\ccmf{{}}{{}}{\assemb}{\ac} \beqv{\gM}
 \ccmf{{}}{{}}{(\ccmf{{}}{{}}{\assemb}{\assemb'})}{\ac}}\;.
\end{equation}

Let $\assemb''$ be the new assembler in executable form obtained by
assembling $\assemb'$ by means of $\assemb$, i.e.\
$\assemb'' = \ccmf{{}}{{}}{\assemb}{\assemb'}$.
Now, $\assemb'$ could be assembled by means of $\assemb''$ instead of
$\assemb$.
In case $\assemb''$ produces more compact executable codes than
$\assemb$, this would result in a new assembler in executable form that
is more compact.
Let $\assemb'''$ be the new assembler in executable form obtained by
assembling $\assemb'$ by means of $\assemb''$, i.e.\
$\assemb''' = \ccmf{{}}{{}}{\assemb''}{\assemb'} =
 \ccmf{{}}{{}}{(\ccmf{{}}{{}}{\assemb}{\assemb'})}{\assemb'}$.
If $\assemb'$ is correct, then $\assemb''$ and $\assemb'''$ produce
the same executable codes.
That is,
\begin{equation}
\label{eqn-equiv-assemblers}
\assemb'' \beqv{\gM} \assemb'''\;.
\end{equation}
This is easy to see: rewriting in terms of $\assemb$ and $\assemb'$
yields
\begin{equation}
\label{eqn-equiv-assemblers-long}
\ccmf{{}}{{}}{\assemb}{\assemb'} \beqv{\gM}
\ccmf{{}}{{}}{(\ccmf{{}}{{}}{\assemb}{\assemb'})}{\assemb'}\;,
\end{equation}
which follows immediately from~(\ref{eqn-correct-assembler}).

Now, $\assemb'$ could be assembled by means of $\assemb'''$ instead of
$\assemb''$.
However, if $\assemb'$ is correct, this would result in $\assemb'''$
again.
That is,
\begin{equation}
\label{eqn-assembler-fix}
\assemb''' = \ccmf{{}}{{}}{\assemb'''}{\assemb'}\;.
\end{equation}
This is easy to see as well: rewriting the left-hand side in terms of
$\assemb'$ and $\assemb''$ yields
\begin{equation}
\label{eqn-assembler-fix-long}
\ccmf{{}}{{}}{\assemb''}{\assemb'} =
\ccmf{{}}{{}}{\assemb'''}{\assemb'}\;,
\end{equation}
which follows immediately from~(\ref{eqn-equiv-assemblers}).
The phenomenon expresses by equation~(\ref{eqn-assembler-fix}) is
called the assembler fixed point.

In theoretical computer science, correctness of a program is taken to
mean that the program satisfies a mathematically precise specification
of it.
For the assembler $\assemb'$,
$\Forall{\ac \in \ACN}{\ccmf{{}}{{}}{\psi(\assemb')}{\ac} = \psi(\ac)}$
would be an obvious mathematically precise specification.
More often than not, practitioners have a more empirical view on the
correctness of a program that is a new program serving as a replacement
for an old one on a specific machine: correctness of the new program is
taken to mean that the old program and the new program give rise to the
same behaviour on that machine.
The correctness criterion for new assemblers given above, as well as
the correctness criterion for new compilers given below, is based on
this empirical view.

\subsection{The Compiler Fixed Point}
\label{subsect-compiler-fix}

In this subsection, we consider the issue of producing a new compiler
for some source code notation using an existing one.
Compilers may produce assembly code, executable code or both.
We deal with the case where compilers produce assembly code only.
The reason for this choice will be explained at the end this subsection.

Let $\gM = \tup{\Bseq,\indfam{\mf_n}{n \in \Nat},\Exec}$ be a code
controlled machine structure, let $\tup{\SCN,\psis}$ be a control
code notation for $\gM$ that belongs to $\SC{\gM}$, and let
$\tup{\ACN,\psia}$ be a control code notation for $\gM$ that belongs to
$\AC{\gM}$.
Suppose that $\transl{\compil}{\SCN}{\ACN}{\ACN}$ is an existing
compiler for $\SCN$ and $\transl{\assemb}{\ACN}{\Exec}{\Exec}$ is an
existing assembler for $\ACN$.
The existing compiler is in assembly form.
However, a compiler in executable form can always be obtained from a
compiler in assembly form by means of the existing assembler.
Suppose further that a new compiler
$\transl{\compil'}{\SCN}{\ACN}{\SCN}$ for $\SCN$ is made available.
This new compiler is not in assembly form.
It needs to be compiled by means of the existing compiler.
The new compiler is considered correct if
\begin{equation}
\label{eqn-correct-compiler}
\begin{array}[c]{@{}l@{}}
\Forall{\src \in \SCN}{{}}
\\ \;\;
 \ccmf{{}}{{}}
  {\assemb}
  {(\ccmf{{}}{{}}{(\ccmf{{}}{{}}{\assemb}{\compil})}{\src})}
\\ \quad {} \beqv{\gM}
 \ccmf{{}}{{}}
  {\assemb}
  {(\ccmf{{}}{{}}
     {(\ccmf{{}}{{}}
        {\assemb}
        {(\ccmf{{}}{{}}
           {(\ccmf{{}}{{}}{\assemb}{\compil})}
           {\compil'})})}
     {\src})}\;.
\end{array}
\end{equation}

Let $\compil''$ be the new compiler in assembly form obtained by
compiling $\compil'$ by means of $\compil$, i.e.\
$\compil'' =
 \ccmf{{}}{{}}{(\ccmf{{}}{{}}{\assemb}{\compil})}{\compil'}$.
Now, $\compil'$ could be compiled by means of $\compil''$ instead of
$\compil$.
In case $\compil''$ produces more compact assembly codes than
$\compil$, this would result in a new compiler in assembly form that
is more compact.
Let $\compil'''$ be the new compiler in assembly form obtained by
compiling $\compil'$ by means of $\compil''$, i.e.\
$\compil''' =
 \ccmf{{}}{{}}{(\ccmf{{}}{{}}{\assemb}{\compil''})}{\compil'} =
 \ccmf{{}}{{}}
  {(\ccmf{{}}{{}}
     {\assemb}
     {(\ccmf{{}}{{}}{(\ccmf{{}}{{}}{\assemb}{\compil})}{\compil'})})}
  {\compil'}$.
If $\compil'$ is correct, then $\compil''$ and $\compil'''$
produce the same assembly codes.
That is,
\begin{equation}
\label{eqn-equiv-compilers}
\ccmf{{}}{{}}{\assemb}{\compil''} \beqv{\gM}
\ccmf{{}}{{}}{\assemb}{\compil'''}\;.
\end{equation}
This is easy to see: rewriting in terms of $\assemb$, $\compil$ and
$\compil'$ yields
\begin{equation}
\label{eqn-equiv-compilers-long}
\begin{array}[c]{@{}l@{}}
\ccmf{{}}{{}}
 {\assemb}
 {(\ccmf{{}}{{}}{(\ccmf{{}}{{}}{\assemb}{\compil})}{\compil'})}
\\ \;\; {} \beqv{\gM}
\ccmf{{}}{{}}
 {\assemb}
 {(\ccmf{{}}{{}}
    {(\ccmf{{}}{{}}
       {\assemb}
       {(\ccmf{{}}{{}}
          {(\ccmf{{}}{{}}{\assemb}{\compil})}
          {\compil'})})}
    {\compil'})}\;,
\end{array}
\end{equation}
which follows immediately from~(\ref{eqn-correct-compiler}).

Now, $\compil'$ could be compiled by means of $\compil'''$ instead
of $\compil''$.
However, if $\compil'$ is correct, this would result in $\compil'''$
again.
That is,
\begin{equation}
\label{eqn-compiler-fix}
\compil''' =
\ccmf{{}}{{}}{(\ccmf{{}}{{}}{\assemb}{\compil'''})}{\compil'}\;.
\end{equation}
This is easy to see as well: rewriting the left-hand side in terms of
$\assemb$, $\compil'$ and $\compil''$ yields
\begin{equation}
\label{eqn-compiler-fix-long}
\ccmf{{}}{{}}{(\ccmf{{}}{{}}{\assemb}{\compil''})}{\compil'} =
\ccmf{{}}{{}}{(\ccmf{{}}{{}}{\assemb}{\compil'''})}{\compil'}\;,
\end{equation}
which follows immediately from~(\ref{eqn-equiv-compilers}).
The phenomenon expresses by equation~(\ref{eqn-compiler-fix}) is
called the compiler fixed point.
It is a non-trivial insight among practitioners involved in matters such
as software configuration and system administration.

The explanation of the compiler fixed point proceeds similar to the
explanation of the assembler fixed point in
Section~\ref{subsect-assembler-fix}, but it is more complicated.
The complication vanishes if compilers that produce executable code are
considered.
In that case, due to the very abstract level at which the issues are
considered, the explanation of the compiler fixed point is essentially
the same as the explanation of the assembler fixed point.

\section{Intermediate Code Notations and Interpreters}
\label{sect-interpreter}

Sometimes, practitioners distinguish additional kinds of control codes.
Intermediate code is a frequently used generic name for those additional
kinds of control codes.
Source code is often implemented by producing executable code for some
code controlled machine by means of a compiler or a compiler and an
assembler.
Sometimes, source code is implemented by means of a compiler and an
interpreter.
In that case, the compiler used produces intermediate code.
The interpreter is a control code corresponding to an executable code of
a code controlled machine that makes that code controlled machine behave
as if it is another code controlled machine controlled by an
intermediate code.

In this section, we briefly consider the issue of the correctness of
such a combination of a compiler and an interpreter.

\subsection{Intermediate Code Notations}
\label{subsect-intermediate}

At the level of control codes for machine structures, like the control
code notations that are to be considered assembly code notations and the
control code notations that are to be considered source code notations,
the control code notations that are to be considered intermediate code
notations of some kind cannot be characterized.
It happens to be sufficient for many issues concerning compilers and
interpreters, including the one considered in this section, to simply
assume that some collection of control code notations comprises the
intermediate code notations of interest.

Henceforth, we assume that, for each machine structure $\gM$, a set
$\IC{\gM}$ of control code notations for $\gM$ has been given.
The members of $\IC{\gM}$ are called \emph{intermediate code notations}
for $\gM$.

The following gives an idea of the grounds on which control code
notations are classified as intermediate code notation.
An intermediate code notation is a control code notation that resembles
an assembly code notation, but it is not specific to any machine.
Often, it is specific to a source code notation or a family of source
code notations.

An intermediate code notation comes into play if source code is
implemented by means of a compiler and an interpreter.
However, compilers for intermediate code notations are found where
interpretation is largely eliminated in favour of just-in-time
compilation, see e.g.~\cite{Ayc03a}, which is material to contemporary
programming languages such as Java and C\#.

In the case where an intermediate code notation is specific to a family
of source code notations, it is a common intermediate code notation for
the source code notations concerned.
The Common Intermediate Language from the .NET Framework~\cite{WHA03a}
is an example of a common intermediate code notation.

\subsection{Interpreters}
\label{subsect-interpreter}

Interpreters are quite different from assemblers and compilers.
An assembler for an assembly code notation makes a code controlled
machine transform members of the assembly code notation into
executable codes and a compiler for a source code notation makes a code
controlled machine transform members of the source code notation into
members of an assembly code notation or executable codes, whereas an
interpreter for an intermediate code notation makes a code controlled
machine behave as if it is a code controlled machine for which the
members of the intermediate code notation serve as executable codes.

We consider the correctness of an interpreter combined with a compiler
going with it.
The correctness criterion given below is in the spirit of the empirical
view on correctness discussed at the end of
Section~\ref{subsect-assembler-fix}.

Let $\gM = \tup{\Bseq,\indfam{\mf_n}{n \in \Nat},\Exec}$ be a code
controlled machine structure,
let $\tup{\SCN,\psis}$ be a control code notation for $\gM$ that
belongs to $\SC{\gM}$,
let $\tup{\ICN,\psii}$ be a control code notation for $\gM$ that
belongs to $\IC{\gM}$, and
let $\tup{\ACN,\psia}$ be a control code notation for $\gM$ that
belongs to $\AC{\gM}$.
Suppose that $\transl{\compila}{\SCN}{\ACN}{\ACN}$ is an existing
compiler for $\SCN$ and $\transl{\assemb}{\ACN}{\Exec}{\Exec}$
is an existing assembler for $\ACN$.
The compiler $\compila$ lets $\gM$ transform source codes into assembly
codes.
Suppose further that a new compiler
$\transl{\compili}{\SCN}{\ICN}{\ACN}$ for $\SCN$ and a new interpreter
$\interp \in \Exec$ for $\ICN$ are made available.
The compiler $\compili$ lets $\gM$ transform source codes into
intermediate codes.

The combination of $\compili$ and $\interp$ is considered correct if
\begin{equation}
\label{eqn-correct-interpreter}
\begin{array}[c]{@{}l@{}}
\Forall
 {\src \in \SCN,\seq{\nm{bs}_1,\ldots,\nm{bs}_m} \in \seqof{\Bseq}}
 {{}}
\\ \;\;
 \ccmf{\gM}{{}}
  {(\ccmf{\gM}{{}}
     {\assemb}
     {(\ccmf{\gM}{{}}
              {(\ccmf{\gM}{{}}{\assemb}{\compila})}
              {\src})})}
  {\nm{bs}_1,\ldots,\nm{bs}_m}
\\ \quad {} =
 \ccmf{\gM}{{}}
  {\interp}
  {(\ccmf{\gM}{{}}{(\ccmf{\gM}{{}}{\assemb}{\compili})}{\src)},
   \nm{bs}_1,\ldots,\nm{bs}_m}\;.
\end{array}
\end{equation}

While being controlled by an interpreter, the behaviour of a code
controlled machine can be looked upon as another code controlled machine
of which the executable codes are the intermediate codes involved.
The latter machine might appropriately be called a virtual machine.
By means of interpreters, the same virtual machine can be obtained on
different machines.
Thus, all machine-de\-pendencies are taken care of by interpreters.
A well-known virtual machine is the Java Virtual Machine~\cite{LY96a}.

\section{Bit Sequence Represented Control Code Notations}
\label{sect-bs-repr-cc}

In order to make a code controlled machine transform members of one
control code notation into members of another control code notation,
like in assembling and compiling, control codes that are not bit
sequences must be represented by bit sequences.
To simplify matters, we assumed up to now that all control code
notations are collections of bit sequences.
In this section, we present the adaptations needed in the preceding
sections when withdrawing this assumption.
It happens that the changes are small.

\subsection*{The Concept of Bit Sequence Represented Control Code
  Notation}

First of all, we have to generalize the concept of control code notation
slightly.

Let $\gM = \tup{\Bseq,\indfam{\mf_n}{n \in \Nat},\Exec}$ be a code
controlled machine structure.
Then a \emph{bit sequence represented control code notation} for $\gM$
consists of a set $\CCN$, a function $\funct{\psi}{\CCN}{\Exec}$, and an
injective function $\funct{\rho}{\CCN}{\Bseq}$.
For all $c \in \CCN$, $\rho(c)$ is called the
\emph{bit sequence representation} of $c$ on $\gM$.
The function $\rho$ is called the \emph{bs-representation function} of
$\CCN$.

Let $\tup{\CCN,\psi,\rho}$ be a bit sequence represented control code
notation for a code controlled machine structure
$\tup{\Bseq,\indfam{\mf_n}{n \in \Nat},\Exec}$.
Then we assume that $\psi(c) = c$ for all $c \in \CCN \inter \Exec$,
$\rho(c') = c'$ for all $c' \in \CCN \inter \Bseq$, and
$\rho(c'') = c''$ for all $c'' \in \CCN$ with $\rho(c'') \in \Exec$.
The last assumption can be paraphrased as follows: if an executable code
is the bit sequence representation of some control code, then it is its
own bit sequence representation.
It excludes bs-representation functions that inadvertently produce
executable codes.

\subsection*{The Special Notation
 $\transl{\nm{cc}}{\CCN'}{\CCN''}{\CCN}$}

We have to change the definition of the special notation
$\transl{\nm{cc}}{\CCN'}{\CCN''}{\CCN}$ slightly.

Let $\gM = \tup{\Bseq,\indfam{\mf_n}{n \in \Nat},\Exec}$ be a code
controlled machine structure, and let $\tup{\CCN,\psi,\rho}$,
$\tup{\CCN',\psi',\rho'}$ and $\tup{\CCN'',\psi'',\rho''}$ be bit
sequence represented control code notations for $\gM$.
Then we write $\transl{\nm{cc}}{\CCN'}{\CCN''}{\CCN}$
for
\begin{ldispl}
\nm{cc} \in \CCN \And
\Forall{\nm{cc}' \in \CCN'}
 {(\Exists{\nm{cc}'' \in \CCN''}
  {\ccmf{\gM}{{}}{\psi(\nm{cc})}{\rho'(\nm{cc}')} =
   \rho''(\nm{cc}'')})}\;.
\end{ldispl}%

\subsection*{The Explanation of the Assembler Fixed Point}

In the explanation of the assembler fixed point given in
Section~\ref{subsect-assembler-fix}, we have to replace the definitions
of $\assemb''$ and $\assemb'''$ by
$\assemb'' = \ccmf{{}}{{}}{\assemb}{\rho(\assemb')}$ and
$\assemb''' =
 \ccmf{{}}{{}}
  {(\ccmf{{}}{{}}{\assemb}{\rho(\assemb')})}
  {\rho(\assemb')}$,
assuming that $\rho$ is the bs-representation function of $\ACN$.
Moreover, we have to adapt Formulas~(\ref{eqn-correct-assembler}),
(\ref{eqn-equiv-assemblers-long}), (\ref{eqn-assembler-fix}), and
(\ref{eqn-assembler-fix-long}) slightly.
Formula~(\ref{eqn-correct-assembler}) must be replaced by
\begin{ldispl}
\Forall{\ac \in \ACN}
{\ccmf{{}}{{}}{\assemb}{\rho(\ac)} \beqv{\gM}
 \ccmf{{}}{{}}{(\ccmf{{}}{{}}{\assemb}{\rho(\assemb')})}{\rho(\ac)}}\;.
\end{ldispl}%
Formula~(\ref{eqn-equiv-assemblers-long}) must be replaced by
\begin{ldispl}
\ccmf{{}}{{}}{\assemb}{\rho(\assemb')} \beqv{\gM}
\ccmf{{}}{{}}
  {(\ccmf{{}}{{}}{\assemb}{\rho(\assemb')})}
  {\rho(\assemb')}\;.
\end{ldispl}%
Formula~(\ref{eqn-assembler-fix}) must be replaced by
\begin{ldispl}
\assemb''' = \ccmf{{}}{{}}{\assemb'''}{\rho(\assemb')}\;.
\end{ldispl}%
Formula~(\ref{eqn-assembler-fix-long}) must be replaced by
\begin{ldispl}
\ccmf{{}}{{}}{\assemb''}{\rho(\assemb')} =
\ccmf{{}}{{}}{\assemb'''}{\rho(\assemb')}\;.
\end{ldispl}%

\subsection*{The Explanation of the Compiler Fixed Point}

In the explanation of the compiler fixed point given in
Section~\ref{subsect-compiler-fix}, we have to replace the definitions
of $\compil''$ and $\compil'''$ by
$\compil'' =
 \ccmf{{}}{{}}
  {(\ccmf{{}}{{}}{\assemb}{\rhoa(\compil)})}
  {\rhos(\compil')}$
and
$\compil''' =
 \ccmf{{}}{{}}
  {(\ccmf{{}}{{}}
     {\assemb}
     {(\ccmf{{}}{{}}
        {(\ccmf{{}}{{}}{\assemb}{\rhoa(\compil)})}
        {\rhos(\compil')})})}
  {\rhos(\compil')}$,
assuming that $\rhos$ is the bs-representation function of $\SCN$ and
$\rhoa$ is the bs-representation function of $\ACN$.
Moreover, we have to adapt Formulas~(\ref{eqn-correct-compiler}),
(\ref{eqn-equiv-compilers-long}), (\ref{eqn-compiler-fix}), and
(\ref{eqn-compiler-fix-long}) slightly.
Formula~(\ref{eqn-correct-compiler}) must be replaced by
\begin{ldispl}
\Forall{\src \in \SCN}{{}}
\\ \;\;
 \ccmf{{}}{{}}
  {\assemb}
  {(\ccmf{{}}{{}}
     {(\ccmf{{}}{{}}{\assemb}{\rhoa(\compil)})}
     {\rhos(\src)})}
\\ \quad {} \beqv{\gM}
 \ccmf{{}}{{}}
  {\assemb}
  {(\ccmf{{}}{{}}
     {(\ccmf{{}}{{}}
        {\assemb}
        {(\ccmf{{}}{{}}
           {(\ccmf{{}}{{}}{\assemb}{\rhoa(\compil)})}
           {\rhos(\compil')})})}
     {\rhos(\src)})}\;.
\end{ldispl}%
Formula~(\ref{eqn-equiv-compilers-long}) must be replaced by
\begin{ldispl}
\ccmf{{}}{{}}
 {\assemb}
 {(\ccmf{{}}{{}}
    {(\ccmf{{}}{{}}{\assemb}{\rhoa(\compil)})}
    {\rhos(\compil')})}
\\ \;\; {} \beqv{\gM}
\ccmf{{}}{{}}
 {\assemb}
 {(\ccmf{{}}{{}}
    {(\ccmf{{}}{{}}
       {\assemb}
       {(\ccmf{{}}{{}}
          {(\ccmf{{}}{{}}{\assemb}{\rhoa(\compil)})}
          {\rhos(\compil')})})}
    {\rhos(\compil')})}\;.
\end{ldispl}%
Formula~(\ref{eqn-compiler-fix}) must be replaced by
\begin{ldispl}
\compil''' =
\ccmf{{}}{{}}{(\ccmf{{}}{{}}{\assemb}{\compil'''})}{\rhos(\compil')}\;.
\end{ldispl}%
Formula~(\ref{eqn-compiler-fix-long}) must be replaced by
\begin{ldispl}
\ccmf{{}}{{}}{(\ccmf{{}}{{}}{\assemb}{\compil''})}{\rhos(\compil')} =
\ccmf{{}}{{}}{(\ccmf{{}}{{}}{\assemb}{\compil'''})}{\rhos(\compil')}\;.
\end{ldispl}%

\subsection*{The Correctness Criterion for Interpreters}

The correctness criterion for interpreters given in
Section~\ref{subsect-interpreter}, i.e.\
Formula~(\ref{eqn-correct-interpreter}), must be replaced by
\begin{ldispl}
\Forall
 {\src \in \SCN,\seq{\nm{bs}_1,\ldots,\nm{bs}_m} \in \seqof{\Bseq}}
 {{}}
\\ \;\;
 \ccmf{\gM}{{}}
  {(\ccmf{\gM}{{}}
     {\assemb}
     {(\ccmf{\gM}{{}}
              {(\ccmf{\gM}{{}}{\assemb}{\rhoa(\compila)})}
              {\rhos(\src)})})}
  {\nm{bs}_1,\ldots,\nm{bs}_m}
\\ \quad {} =
 \ccmf{\gM}{{}}
  {\interp}
  {(\ccmf{\gM}{{}}
     {(\ccmf{\gM}{{}}{\assemb}{\rhoa(\compili)})}
     {\rhos(\src))},
   \nm{bs}_1,\ldots,\nm{bs}_m}\;,
\end{ldispl}%
assuming that $\rhos$ is the bs-representation function of $\SCN$ and
$\rhoa$ is the bs-representation function of $\ACN$.

\section{An Execution Architecture for Machine Structures}
\label{sect-exearch}

Machine structures in themselves are not always sufficient to explain
issues concerning control codes that are independent of the details of
the behaviours that are controlled.
In cases where systems that provide execution environments for the
executable codes of machine structures are involved, such as in the case
of portability of control codes, an abstract model of such systems is
needed.
In this section, we outline an appropriate model.
This model is referred to as the execution architecture for code
controlled machine structures.
It is a synthetic execution architecture in the sense of~\cite{BP04a}.
It can be looked upon as an abstract model of operating systems
restricted to file management facilities and facilities for loading and
execution of executable codes.

The execution architecture for code controlled machine structures, which
is parameterized by a code controlled machine structure $\gM$, is an
abstract model of a system that provides an execution environment for
the executable codes of $\gM$.
It can be looked upon as a machine.
This machine is operated by means of instructions that either yield a
reply or diverge.
The possible replies are $\True$ and $\False$.
File names are used in the instructions to refer to the bit sequences
present in the machine.
It is assumed that a countably infinite set $\Fnm$ of \emph{file names}
has been given.
While designing the instruction set, we focussed on convenience of use
rather than minimality.

Let $\gM = \tup{\Bseq,\indfam{\mf_n}{n \in \Nat},\Exec}$ be a code
controlled machine structure.
Then the instruction set consists of the following instructions:
\begin{iteml}
\item
for each $\fnm \in \Fnm$ and $\bs \in \Bseq$,
a \emph{set} instruction $\seti{:}\fnm{:}\bs$;
\item
for each $\fnm \in \Fnm$,
a \emph{remove} instruction $\rmvi{:}\fnm$;
\item
for each $\fnm_1,\fnm_2 \in \Fnm$,
a \emph{copy} instruction $\copi{:}\fnm_1{:}\fnm_2$;
\item
for each $\fnm_1,\fnm_2 \in \Fnm$,
a \emph{move} instruction $\movi{:}\fnm_1{:}\fnm_2$;
\item
for each  $\fnm_1,\fnm_2 \in \Fnm$,
a \emph{concatenation} instruction $\cati{:}\fnm_1{:}\fnm_2$;
\item
for each $\fnm_1,\fnm_2 \in \Fnm$,
a \emph{test on equality} instruction $\equi{:}\fnm_1{:}\fnm_2$;
\item
for each $\fnm_1,\fnm_2 \in \Fnm$,
a \emph{test on difference} instruction $\neqi{:}\fnm_1{:}\fnm_2$;
\item
for each $\fnm \in \Fnm$,
a \emph{test on existence} instruction $\exii{:}\fnm$;
\item
for each $\fnm \in \Fnm$,
a \emph{load} instruction $\loai{:}\fnm$;
\item
for each $\fnm_1,\ldots,\fnm_m,\fnm'_1,\ldots,\fnm'_n \in \Fnm$,
an \emph{execute} instruction
$\exei{:}\fnm_1{:}\ldots{:}\fnm_m\iosep
         \fnm'_1{:}\ldots{:}\fnm'_n$.
\end{iteml}
We write $\Instr$ for this instruction set.

We say that a file name is in use if it has a bit sequence assigned.
A state of the machine comprises the file names that are in use, the bit
sequences assigned to those file names, a flag indicating whether there
is a loaded executable code, and the loaded executable code if there is
one.

The instructions can be explained in terms of the effect that they have
and the reply that they yield as follows:
\begin{iteml}
\item
$\seti{:}\fnm{:}\bs$: the file name $\fnm$ is added to the file names in
use if it is not in use, the bit sequence $\bs$ is assigned to $\fnm$,
and the reply is $\True$;
\item
$\rmvi{:}\fnm$: if the file name $\fnm$ is in use, then it is removed
from the file names in use and the reply is $\True$; otherwise, nothing
changes and the reply is $\False$;
\item
$\copi{:}\fnm_1{:}\fnm_2$: if the file name $\fnm_1$ is in use, then the
file name $\fnm_2$ is added to the file names in use if it is not in
use, the bit sequence assigned to $\fnm_1$ is assigned to $\fnm_2$, and
the reply is $\True$; otherwise, nothing changes and the reply is
$\False$;
\item
$\movi{:}\fnm_1{:}\fnm_2$: if the file name $\fnm_1$ is in use, then the
file name $\fnm_2$ is added to the file names in use if it is not in
use, the bit sequence assigned to $\fnm_1$ is assigned to $\fnm_2$,
$\fnm_1$ is removed from the file names in use, and the reply is
$\True$; otherwise, nothing changes and the reply is $\False$;
\item
$\cati{:}\fnm_1{:}\fnm_2$: if the file names $\fnm_1$ and $\fnm_2$ are
in use, then the concatenation of the bit sequence assigned to $\fnm_2$
and the bit sequence assigned to $\fnm_1$ is assigned to $\fnm_2$ and
the reply is $\True$; otherwise, nothing changes and the reply is
$\False$;
\item
$\equi{:}\fnm_1{:}\fnm_2$: if the file names $\fnm_1$ and $\fnm_2$ are
in use and the bit sequence assigned to $\fnm_1$ equals the bit sequence
assigned to $\fnm_2$, then nothing changes and the reply is $\True$;
otherwise, nothing changes and the reply is $\False$;
\item
$\neqi{:}\fnm_1{:}\fnm_2$: if the file names $\fnm_1$ and $\fnm_2$ are
in use and the bit sequence assigned to $\fnm_1$ does not equal the bit
sequence assigned to $\fnm_2$, then nothing changes and the reply is
$\True$; otherwise, nothing changes and the reply is $\False$;
\item
$\exii{:}\fnm$: if the file name $\fnm$ is in use, then nothing changes
and the reply is $\True$; otherwise, nothing changes and the reply is
$\False$;
\item
$\loai{:}\fnm$: if the file name $\fnm$ is in use and the bit sequence
assigned to $\fnm$ is a member of $\Exec$, then the bit sequence
assigned to $\fnm$ is loaded and the reply is $\True$; otherwise,
nothing changes and the reply is $\False$;
\item
$\exei{:}\fnm_1{:}\ldots{:}\fnm_m\iosep\fnm'_1{:}\ldots{:}\fnm'_n$:
if the file names $\fnm_1,\ldots,\fnm_m$ have bit sequences assigned,
say $\bs_1,\ldots,\bs_m$, and there is a loaded executable code, say
$x$, then:
\begin{iteml}
\item
if $\ccmf{\gM}{1}{x}{\bs_1,\ldots,\bs_m} \in \Bseq$, then:
\begin{iteml}
\item
$\ccmf{\gM}{i}{x}{\bs_1,\ldots,\bs_m}$ is assigned to $\fnm'_i$ for each
$i$ with $1 \leq i \leq n$ such that
$\ccmf{\gM}{i}{x}{\bs_1,\ldots,\bs_m} \in \Bseq$,
\item
$\fnm'_i$ is removed from the file names in use for each $i$ with
$1 \leq i \leq n$ such that
$\ccmf{\gM}{i}{x}{\bs_1,\ldots,\bs_m} = \Mea$,
\end{iteml}
and the reply is $\True$;
\item
if $\ccmf{\gM}{1}{x}{\bs_1,\ldots,\bs_m} = \Mea$, then nothing changes
and the reply is $\False$;
\item
if $\ccmf{\gM}{1}{x}{\bs_1,\ldots,\bs_m} = \Div$, then the machine does
not halt;
\end{iteml}
otherwise, nothing changes and the reply is $\False$.
\end{iteml}
Note that there are three cases in which the instruction
$\exei{:}\fnm_1{:}\ldots{:}\fnm_m\iosep\fnm'_1{:}\ldots{:}\fnm'_n$
yields the reply $\False$:
(a)~there is no loaded executable code;
(b)~there is some file name among $\fnm_1,\ldots,\fnm_m$ that is not in
use;
(c)~there is no output produced, although the machine halts.

The instructions of which the effect depends on the code controlled
machine structure $\gM$ are the load and execute instructions only.
All other instructions could be eliminated in favour of executable
codes, assigned to known file names.
However, we believe that elimination of these instructions would not
contribute to a useful execution architecture.
The distinction made between loading and execution of executable codes
allows for telling load-time errors from run-time errors.

\section{Thread Algebra}
\label{sect-TA}

The execution architecture for code controlled machine structures
outlined above can be looked upon as a machine which is operated by
means of instructions that yield $\True$ or $\False$ as reply.
In cases where this execution architecture is needed to explain issues
concerning control codes, such as in the case of portability of control
codes, processes that operate upon the execution architecture have to be
described.
An existing extension of \BTA\ (Basic Thread Algebra), first presented
in~\cite{BP02a}, is tailored to the description of processes that
operate upon machines of the kind to which the execution architecture
belongs.
Therefore, we have chosen to use in Section~\ref{sect-cc-exearch} the
extension of \BTA\ in question to describe processes that operate upon
the execution architecture.
In this section, we review \BTA, including guarded recursion and the
approximation induction principle, and the relevant extension.

\subsection{Basic Thread Algebra}
\label{subsect-BTA}

\BTA\ is concerned with the behaviours produced by deterministic sequential
programs under execution.
The behaviours concerned are called \emph{threads}.
It does not matter how programs are executed: threads may originate from
execution by a computer, or they may originate from execution by a human
operator.
In~\cite{BL02a}, \BTA\ is introduced under the name \BPPA\ (Basic
Polarized Process Algebra).

In \BTA, it is assumed that there is a fixed but arbitrary set of
\emph{basic actions}~$\BAct$.
The intuition is that each basic action performed by a thread is taken
as a command to be processed by a service provided by the execution
environment of the thread.
The processing of a command may involve a change of state of the
service concerned.
At completion of the processing of the command, the service produces a
reply value.
This reply is either $\True$ or $\False$ and is returned to the thread
concerned.

Although \BTA\ is one-sorted, we make this sort explicit.
The reason for this is that we will extend \BTA\ with additional sorts
in Section~\ref{subsect-apply}.

The algebraic theory \BTA\ has one sort: the sort $\Thr$ of
\emph{threads}.
\BTA\ has the following constants and operators:
\begin{iteml}
\item
the \emph{deadlock} constant $\const{\Dead}{\Thr}$;
\item
the \emph{termination} constant $\const{\Stop}{\Thr}$;
\item
for each $a \in \BAct$, the binary \emph{postconditional composition}
operator $\funct{\pcc{\ph}{a}{\ph}}{\Thr \x \Thr}{\Thr}$.
\end{iteml}
Terms of sort $\Thr$ are built as usual.
Throughout the paper, we assume that there are infinitely many variables
of sort $\Thr$, including $u,v,w$.

We use infix notation for postconditional composition.
We introduce \emph{action prefixing} as an abbreviation: $a \bapf p$,
where $p$ is a term  of sort $\Thr$, abbreviates $\pcc{p}{a}{p}$.

Let $p$ and $q$ be closed terms of sort $\Thr$ and $a \in \BAct$.
Then $\pcc{p}{a}{q}$ will perform action $a$, and after that proceed as
$p$ if the processing of $a$ leads to the reply $\True$ (called a
positive reply) and proceed as $q$ if the processing of $a$ leads to the
reply $\False$ (called a negative reply).

Each closed term of sort $\Thr$ from the language of \BTA\ denotes a
finite thread, i.e.\ a thread of which the length of the sequences of
actions that it can perform is bounded.
Guarded recursive specifications give rise to infinite threads.

A \emph{guarded recursive specification} over \BTA\ is a set of
recursion equations $E = \set{X = t_X \where X \in V}$, where $V$ is a
set of variables of sort $\Thr$ and each $t_X$ is a term of sort $\Thr$
that has the form $\Dead$, $\Stop$ or $\pcc{t}{a}{t'}$.
We write $\vars(E)$ for the set of all variables that occur on the
left-hand side of an equation in $E$.
We are only interested in models of \BTA\ in which guarded recursive
specifications have unique solutions, such as the projective limit model
of \BTA\ presented in~\cite{BB03a}.

We extend \BTA\ with guarded recursion by adding constants for solutions
of guarded recursive specifications and axioms concerning these
additional constants.
For each guarded recursive specification $E$ and each $X \in \vars(E)$,
we add a constant of sort $\Thr$ standing for the unique solution of $E$
for $X$ to the constants of \BTA.
The constant standing for the unique solution of $E$ for $X$ is denoted
by $\rec{X}{E}$.
Moreover, we add the axioms for guarded recursion given in
Table~\ref{axioms-rec} to \BTA,%
\begin{table}[!t]
\caption{Axioms for guarded recursion}
\label{axioms-rec}
\begin{eqntbl}
\begin{saxcol}
\rec{X}{E} = \rec{t_X}{E} & \mif X \!=\! t_X \in E & \axiom{RDP}
\\
E \Implies X = \rec{X}{E} & \mif X \in \vars(E)    & \axiom{RSP}
\end{saxcol}
\end{eqntbl}
\end{table}
where we write $\rec{t_X}{E}$ for $t_X$ with, for all $Y \in \vars(E)$,
all occurrences of $Y$ in $t_X$ replaced by $\rec{Y}{E}$.
In this table, $X$, $t_X$ and $E$ stand for an arbitrary variable of
sort $\Thr$, an arbitrary term of sort $\Thr$ from the language of \BTA,
and an arbitrary guarded recursive specification over \BTA,
respectively.
Side conditions are added to restrict the variables, terms and guarded
recursive specifications for which $X$, $t_X$ and $E$ stand.
The equations $\rec{X}{E} = \rec{t_X}{E}$ for a fixed $E$ express that
the constants $\rec{X}{E}$ make up a solution of $E$.
The conditional equations $E \Implies X = \rec{X}{E}$ express that this
solution is the only one.

We will write \BTA+\REC\ for \BTA\ extended with the constants for
solutions of guarded recursive specifications and axioms RDP and RSP.

In~\cite{BM05c}, we show that the processes considered in \BTA+\REC\ can
be viewed as processes that are definable over ACP~\cite{Fok00}.

Closed terms of sort $\Thr$ from the language of \BTA+\REC\ that denote
the same infinite thread cannot always be proved equal by means of the
axioms of \BTA+\REC. \sloppy
We introduce the approximation induction principle to remedy this.
The approximation induction principle, \AIP\ in short, is based on the
view that two threads are identical if their approximations up to any
finite depth are identical.
The approximation up to depth $n$ of a thread is obtained by cutting it
off after performing a sequence of actions of length $n$.

\AIP\ is the infinitary conditional equation given in
Table~\ref{axioms-AIP}.%
\begin{table}[!t]
\caption{Approximation induction principle}
\label{axioms-AIP}
\begin{eqntbl}
\begin{axcol}
\AND{n \geq 0} \proj{n}{u} = \proj{n}{v} \Implies u = v   & \axiom{AIP}
\end{axcol}
\end{eqntbl}
\end{table}
Here, following~\cite{BL02a}, approximation of depth $n$ is phrased in
terms of a unary \emph{projection} operator $\projop{n}$.
The axioms for the projection operators are given in
Table~\ref{axioms-pin}.%
\begin{table}[!t]
\caption{Axioms for projection operators}
\label{axioms-pin}
\begin{eqntbl}
\begin{axcol}
\proj{0}{u} = \Dead                                      & \axiom{P0} \\
\proj{n+1}{\Stop} = \Stop                                & \axiom{P1} \\
\proj{n+1}{\Dead} = \Dead                                & \axiom{P2} \\
\proj{n+1}{\pcc{u}{a}{v}} =
                       \pcc{\proj{n}{u}}{a}{\proj{n}{v}} & \axiom{P3}
\end{axcol}
\end{eqntbl}
\end{table}
In this table, $a$ stands for an arbitrary member of $\BAct$.

\subsection{Applying Threads to Services}
\label{subsect-apply}

We extend \BTA+\REC\ to a theory that covers the effects of applying
threads to services.

It is assumed that there is a fixed but arbitrary set of \emph{foci}
$\Foci$ and a fixed but arbitrary set of \emph{methods} $\Meth$.
For  the set of basic actions $\BAct$, we take the set
$\FM = \set{f.m \where f \in \Foci, m \in \Meth}$.
Each focus plays the role of a name of a service provided by the
execution environment that can be requested to process a command.
Each method plays the role of a command proper.
Performing a basic action $f.m$ is taken as making a request to the
service named $f$ to process the command $m$.

We introduce a second sort: the sort $\Srv$ of \emph{services}.
However, we will not introduce constants and operators to build terms
of this sort.
$\Srv$ is a parameter of theories with thread-to-service application.
$\Srv$ is considered to stand for the set of all services.
It is assumed that each service can be represented by a function
$\funct{H}{\neseqof{\Meth}}{\set{\True,\False,\Div}}$ with the
property that
$H(\gamma) = \Div \Implies H(\gamma \conc \seq{m}) = \Div$ for
all $\gamma \in \neseqof{\Meth}$ and $m \in \Meth$.
This function is called the \emph{reply} function of the service.
Given a reply function $H$ and a method $m \in \Meth$, the
\emph{derived} reply function of $H$ after processing $m$, written
$\derive{m}H$, is defined by
$\derive{m}H(\gamma) = H(\seq{m} \conc \gamma)$.

The connection between a reply function $H$ and the service represented
by it can be understood as follows:
\begin{iteml}
\item
if $H(\seq{m}) = \True$, the request to process command $m$ is accepted
by the service, the reply is positive and the service proceeds as
$\derive{m}H$;
\item
if $H(\seq{m}) = \False$, the request to process command $m$ is accepted
by the service, the reply is negative and the service proceeds as
$\derive{m}H$;
\item
if $H(\seq{m}) = \Div$, either the processing of command $m$ by the
service does not halt or the processing of a previous command by the
service did not halt.
\end{iteml}
Henceforth, we will identify a reply function with the service
represented by it.

It is assumed that there is an \emph{undefined service} $\undef$ with
the property that $\undef(\gamma) = \Div$ for all
$\gamma \in \neseqof{\Meth}$.

For each $f \in \Foci$, we introduce the binary \emph{apply} operator
$\funct{\apply{\ph}{f}{\ph}}{\Thr \x \Srv}{\Thr}$.
Intuitively, $\apply{p}{f}{H}$ is the service that evolves from $H$ on
processing all basic actions performed by thread $p$ that are of the
form $f.m$ by $H$.
When a basic action $f.m$ performed by thread $p$ is processed by $H$,
$p$ proceeds on the basis of the reply value produced.

The axioms for the apply operators are given in
Table~\ref{axioms-apply}.%
\begin{table}[!t]
\caption{Axioms for apply}
\label{axioms-apply}
\begin{eqntbl}
\begin{saxcol}
\apply{u}{f}{\undef} = \undef                        & & \axiom{TSA0} \\
\apply{\Stop}{f}{H} = H                              & & \axiom{TSA1} \\
\apply{\Dead}{f}{H} = \undef                         & & \axiom{TSA2} \\
\apply{(\pcc{u}{g.m}{v})}{f}{H} = \undef
                                       & \mif f \neq g & \axiom{TSA3} \\
\apply{(\pcc{u}{f.m}{v})}{f}{H} = \apply{u}{f}{\derive{m}H}
                          & \mif H(\seq{m}) = \True    & \axiom{TSA4} \\
\apply{(\pcc{u}{f.m}{v})}{f}{H} = \apply{v}{f}{\derive{m}H}
                          & \mif H(\seq{m}) = \False   & \axiom{TSA5} \\
\apply{(\pcc{u}{f.m}{v})}{f}{H} = \undef
                          & \mif H(\seq{m}) = \Div     & \axiom{TSA6} \\
(\AND{n \geq 0} \apply{\proj{n}{u}}{f}{H} = \undef) \Implies
\apply{u}{f}{H} = \undef                             & & \axiom{TSA7}
\end{saxcol}
\end{eqntbl}
\end{table}
In this table, $f$ and $g$ stand for arbitrary foci from $\Foci$ and $m$
stands for an arbitrary method from $\Meth$.
The axioms show that $\apply{p}{f}{H}$ does not equal $\undef$ only if
thread $p$ performs no other basic actions than ones of the form $f.m$
and eventually terminates successfully.

Let $p$ be a closed term of sort $\Thr$ from the language of \BTA+\REC\
and $H$ be a closed term of sort $\Srv$.
Then $p$ \emph{converges} from $H$ on $f$ if there exists an
$n \in \Nat$ such that $\apply{\proj{n}{p}}{f}{H} \neq \undef$.
Notice that axiom TSA7 can be read as follows:
if $u$ does not converge from $H$ on $f$, then $\apply{u}{f}{H}$ equals
$\undef$.

The extension of \BTA\ introduced above originates from~\cite{BP02a}.
In the remainder of this paper, we will use just one focus.
We have introduced the general case here because the use of several foci
might be needed on further elaboration of the work presented in this
paper.

\section{The Execution Architecture Services}
\label{sect-exearch-formal}

In order to be able to use the extension of \BTA\ presented above to
describe processes that operate upon the execution architecture for code
controlled machine structures outlined in Section~\ref{sect-exearch}, we
have to associate a service with each state of the execution
architecture.
In this section, we first formalize the execution architecture for code
controlled machine structures and then associate a service with each of
its states.

\subsection{The Execution Architecture Formalized}
\label{subsect-exearch-formal}

The execution architecture for code controlled machine structures
consists of an instruction set, a state set, an effect function, and a
yield function.
The effect and yield functions give, for each instruction $u$ and state
$s$, the state and reply, respectively, that result from processing $u$
in state $s$.

It is assumed that
$\sdiv \not\in
 (\bigcup_{F \in \fsetof{(\Fnm)}} (F \to \BitSeq)) \x
 (\BitSeq \union \set{\Mea})$.
Here, $\sdiv$ stands for a state of divergence.

Let $\gM = \tup{\Bseq,\indfam{\mf_n}{n \in \Nat},\Exec}$ be a code
controlled machine structure.
Then the \emph{execution architecture} for $\gM$ consists of
\begin{iteml}
\item
the instruction set $\Instr$ defined in Section~\ref{sect-exearch};
\item
the state set $S$ defined by
\begin{ldispl}
S =
 \biggl(\biggl(\Union{F \in \fsetof{(\Fnm)}} (F \to \Bseq)\biggr) \x
                (\Exec \union \set{\Mea})\biggr) \union \set{\sdiv}\;;
\end{ldispl}%
\item
the effect function $\funct{\eff}{\Instr \x S}{S}$ defined in
Table~\ref{eff-exearch};%
%
\begin{table}[!t]
\caption{Effect function for an execution architecture
 ($i \in \Instr$)}
\label{eff-exearch}
\begin{eqntbl}
\begin{axcol}
\eff(\seti{:}\fnm{:}\bs,\tup{\sigma,x}) =
\tup{\sigma \owr \maplet{\fnm}{\bs},x}
\\
\eff(\rmvi{:}\fnm,\tup{\sigma,x}) = \tup{\sigma \dsub \set{\fnm},x}
\\
\eff(\copi{:}\fnm_1{:}\fnm_2,\tup{\sigma,x}) =
\tup{\sigma \owr \maplet{\fnm_2}{\sigma(\fnm_1)},x}
 & \mif \fnm_1 \in \dom(\sigma)
\\
\eff(\copi{:}\fnm_1{:}\fnm_2,\tup{\sigma,x}) = \tup{\sigma,x}
 & \mif \fnm_1 \not\in \dom(\sigma)
\\
\eff(\movi{:}\fnm_1{:}\fnm_2,\tup{\sigma,x}) =
\tup{(\sigma \owr \maplet{\fnm_2}{\sigma(\fnm_1)}) \dsub \set{\fnm_1},x}
 & \mif \fnm_1 \in \dom(\sigma)
\\
\eff(\movi{:}\fnm_1{:}\fnm_2,\tup{\sigma,x}) = \tup{\sigma,x}
 & \mif \fnm_1 \not\in \dom(\sigma)
\\
\eff(\cati{:}\fnm_1{:}\fnm_2,\tup{\sigma,x}) =
\tup{\sigma \owr \maplet{\fnm_2}{\sigma(\fnm_2) \conc \sigma(\fnm_1)},x}
 & \mif \fnm_1 \in \dom(\sigma) \And \fnm_2 \in \dom(\sigma)
\\
\eff(\cati{:}\fnm_1{:} \fnm_2,\tup{\sigma,x}) = \tup{\sigma,x}
 & \mif \fnm_1 \not\in \dom(\sigma) \Or \fnm_2 \not\in \dom(\sigma)
\\
\eff(\equi{:}\fnm_1{:}\fnm_2,\tup{\sigma,x}) = \tup{\sigma,x}
\\
\eff(\neqi{:}\fnm_1{:}\fnm_2,\tup{\sigma,x}) = \tup{\sigma,x}
\\
\eff(\exii{:}\fnm,\tup{\sigma,x}) = \tup{\sigma,x}
\\
\eff(\loai{:}\fnm,\tup{\sigma,x}) = \tup{\sigma,\sigma(\fnm)}
 & \mif \fnm \in \dom(\sigma) \And \sigma(\fnm) \in \Exec
\\
\eff(\loai{:}\fnm,\tup{\sigma,x}) = \tup{\sigma,x}
 & \mif \fnm \not\in \dom(\sigma) \Or \sigma(\fnm) \not\in \Exec
\\
\multicolumn{2}{@{}l@{}}
 {\eff(\exei{:}\fnm_1{:}\ldots{:}\fnm_m\iosep\fnm'_1{:}\ldots{:}\fnm'_n,
       \tup{\sigma,x}) =
  \tup{(\ldots (\sigma \owr \sigma'_1) \ldots \owr \sigma'_n),x}} \\
\multicolumn{2}{@{\qquad\;\;}l@{}}
 {\mathsf{where}\;\;
  \begin{array}[t]{@{}l@{\;\;}l@{}}
  \sigma'_i =
  \maplet
   {\fnm'_i}
   {\ccmf{\gM}{i}{x}{\sigma(\fnm_1),\ldots,\sigma(\fnm_m)}}
   & \mif\;
  \ccmf{\gM}{i}{x}{\sigma(\fnm_1),\ldots,\sigma(\fnm_m)} \in \Bseq
  \\
  \sigma'_i = \emptymap
   & \mif\;
  \ccmf{\gM}{i}{x}{\sigma(\fnm_1),\ldots,\sigma(\fnm_m)} = \Mea
  \end{array}
 } \\
\multicolumn{2}{@{\qquad\;\;}l@{}}
 {\mif x \in \Exec \And
       \fnm_1 \in \dom(\sigma) \And \ldots \And
       \fnm_m \in \dom(\sigma) \And
       \ccmf{\gM}{1}{x}{\sigma(\fnm_1),\ldots,\sigma(\fnm_m)} \in \Bseq}
\\
\eff(\exei{:}\fnm_1{:}\ldots{:}\fnm_m\iosep\fnm'_1{:}\ldots{:}\fnm'_n,
     \tup{\sigma,x}) =
\tup{\sigma,x} \\
\multicolumn{2}{@{\qquad\;\;}l@{}}
 {\mif x \not\in \Exec \Or
       \fnm_1 \not\in \dom(\sigma) \Or \ldots \Or
       \fnm_m \not\in \dom(\sigma) \Or
       \ccmf{\gM}{1}{x}{\sigma(\fnm_1),\ldots,\sigma(\fnm_m)} = \Mea}
\\
\eff(\exei{:}\fnm_1{:}\ldots{:}\fnm_m\iosep\fnm'_1{:}\ldots{:}\fnm'_n,
     \tup{\sigma,x}) =
\sdiv \\
\multicolumn{2}{@{\qquad\;\;}l@{}}
 {\mif x \not\in \Exec \Or
       \fnm_1 \not\in \dom(\sigma) \Or \ldots \Or
       \fnm_m \not\in \dom(\sigma) \Or
       \ccmf{\gM}{1}{x}{\sigma(\fnm_1),\ldots,\sigma(\fnm_m)} = \Div}
\\
\eff(i,\sdiv) = \sdiv
\end{axcol}
\end{eqntbl}
\end{table}
\item
the yield function
$\funct{\yld}{\Instr \x S}{\set{\True,\False,\Div}}$ defined in
Table~\ref{yld-exearch}.%
\begin{table}[!t]
\caption{Yield function for an execution architecture
 ($i \in \Instr$)}
\label{yld-exearch}
\begin{eqntbl}
\begin{axcol}
\yld(\seti{:}\fnm{:}\bs,\tup{\sigma,x}) = \True
\\
\yld(\rmvi{:}\fnm,\tup{\sigma,x}) = \True
 & \mif \fnm \in \dom(\sigma)
\\
\yld(\rmvi{:}\fnm,\tup{\sigma,x}) = \False
 & \mif \fnm \not\in \dom(\sigma)
\\
\yld(\copi{:}\fnm_1{:}\fnm_2,\tup{\sigma,x}) = \True
 & \mif \fnm_1 \in \dom(\sigma)
\\
\yld(\copi{:}\fnm_1{:}\fnm_2,\tup{\sigma,x}) = \False
 & \mif \fnm_1 \not\in \dom(\sigma)
\\
\yld(\movi{:}\fnm_1{:}\fnm_2,\tup{\sigma,x}) = \True
 & \mif \fnm_1 \in \dom(\sigma)
\\
\yld(\movi{:}\fnm_1{:}\fnm_2,\tup{\sigma,x}) = \False
 & \mif \fnm_1 \not\in \dom(\sigma)
\\
\yld(\cati{:}\fnm_1{:}\fnm_2,\tup{\sigma,x}) = \True
 & \mif \fnm_1 \in \dom(\sigma) \And \fnm_2 \in \dom(\sigma)
\\
\yld(\cati{:}\fnm_1{:}\fnm_2,\tup{\sigma,x}) = \False
 & \mif \fnm_1 \not\in \dom(\sigma) \Or \fnm_2 \not\in \dom(\sigma)
\\
\yld(\equi{:}\fnm_1{:}\fnm_2,\tup{\sigma,x}) = \True
 & \mif \fnm_1 \in \dom(\sigma) \And \fnm_2 \in \dom(\sigma) \And
        \sigma(\fnm_1) = \sigma(\fnm_2)
\\
\yld(\equi{:}\fnm_1{:}\fnm_2,\tup{\sigma,x}) = \False
 & \mif \fnm_1 \not\in \dom(\sigma) \Or \fnm_2 \not\in \dom(\sigma) \Or
        \sigma(\fnm_1) \neq \sigma(\fnm_2)
\\
\yld(\neqi{:}\fnm_1{:}\fnm_2,\tup{\sigma,x}) = \True
 & \mif \fnm_1 \in \dom(\sigma) \And \fnm_2 \in \dom(\sigma) \And
        \sigma(\fnm_1) \neq \sigma(\fnm_2)
\\
\yld(\neqi{:}\fnm_1{:}\fnm_2,\tup{\sigma,x}) = \False
 & \mif \fnm_1 \not\in \dom(\sigma) \Or \fnm_2 \not\in \dom(\sigma) \Or
        \sigma(\fnm_1) = \sigma(\fnm_2)
\\
\yld(\exii{:}\fnm,\tup{\sigma,x}) = \True
 & \mif \fnm \in \dom(\sigma)
\\
\yld(\exii{:}\fnm,\tup{\sigma,x}) = \False
 & \mif \fnm \not\in \dom(\sigma)
\\
\yld(\loai{:}\fnm,\tup{\sigma,x}) = \True
 & \mif \fnm \in \dom(\sigma) \And \sigma(\fnm) \in \Exec
\\
\yld(\loai{:}\fnm,\tup{\sigma,x}) = \False
 & \mif \fnm \not\in \dom(\sigma) \Or \sigma(\fnm) \not\in \Exec
\\
\multicolumn{2}{@{}l@{}}
 {\yld(\exei{:}\fnm_1{:}\ldots{:}\fnm_m\iosep\fnm'_1{:}\ldots{:}\fnm'_n,
       \tup{\sigma,x}) =
  \True} \\
\multicolumn{2}{@{\qquad\;\;}l@{}}
 {\mif x \in \Exec \And
       \fnm_1 \in \dom(\sigma) \And \ldots \And
       \fnm_m \in \dom(\sigma) \And
       \ccmf{\gM}{1}{x}{\sigma(\fnm_1),\ldots,\sigma(\fnm_m)} \in \Bseq}
\\
\multicolumn{2}{@{}l@{}}
 {\yld(\exei{:}\fnm_1{:}\ldots{:}\fnm_m\iosep\fnm'_1{:}\ldots{:}\fnm'_n,
       \tup{\sigma,x}) =
  \False} \\
\multicolumn{2}{@{\qquad\;\;}l@{}}
 {\mif x \not\in \Exec \Or
       \fnm_1 \not\in \dom(\sigma) \Or \ldots \Or
       \fnm_m \not\in \dom(\sigma) \Or
       \ccmf{\gM}{1}{x}{\sigma(\fnm_1),\ldots,\sigma(\fnm_m)} = \Mea}
\\
\multicolumn{2}{@{}l@{}}
 {\yld(\exei{:}\fnm_1{:}\ldots{:}\fnm_m\iosep\fnm'_1{:}\ldots{:}\fnm'_n,
       \tup{\sigma,x}) =
  \Div} \\
\multicolumn{2}{@{\qquad\;\;}l@{}}
 {\mif x \not\in \Exec \Or
       \fnm_1 \not\in \dom(\sigma) \Or \ldots \Or
       \fnm_m \not\in \dom(\sigma) \Or
       \ccmf{\gM}{1}{x}{\sigma(\fnm_1),\ldots,\sigma(\fnm_m)} = \Div}
\\
\yld(i,\sdiv) = \Div
\end{axcol}
\end{eqntbl}
\end{table}
\end{iteml}
We use the following notation for functions:
$\emptymap$ for the empty function;
$\maplet{d}{r}$ for the function $f$ with $\dom(f) = \set{d}$ such that
$f(d) = r$;
$f \owr g$ for the function $h$ with $\dom(h) = \dom(f) \union \dom(g)$
such that for all $d \in \dom(h)$,\, $h(d) = f(d)$ if
$d \not\in \dom(g)$ and $h(d) = g(d)$ otherwise;
and $f \dsub D$ for the function $g$ with $\dom(g) = \dom(f) \diff D$
such that for all $d \in \dom(g)$,\, $g(d) = f(d)$.

Let $\tup{\sigma,x} \in S$, and let $\fnm \in \Fnm$.
Then $\fnm$ is in use if $\fnm \in \dom(\sigma)$, and there is a loaded
executable code if $x \neq \Mea$.
If $\fnm$ is in use, then $\sigma(\fnm)$ is the bit sequence assigned
to $\fnm$.
If there is a loaded executable code, then $x$ is the loaded executable
code.

Execute instructions can diverge.
When an instruction diverges, a situation arises in which no reply can
be produced and no further instructions can be processed.
This is modelled by $\eff$ producing $\sdiv$ and $\yld$ producing
$\Div$.

\subsection{The Family of Execution Architecture Services}
\label{subsect-services}

Each state of the execution architecture for code controlled machine
structures can be looked upon as a service by assuming that
$\Instr \subseteq \Meth$ and extending the functions $\eff$ and $\yld$
from $\Instr$ to $\Meth$ by stipulating that $\eff(m,s) = \sdiv$ and
$\yld(m,s) = \Div$ for all $m \in \Meth \diff \Instr$ and $s \in S$.

We define, for each $s \in S$, a cumulative effect function
$\funct{\ceff_s}{\seqof{\Meth}}{S}$
in terms of $s$ and $\eff$ as follows:
\begin{ldispl}
\ceff_s(\emptyseq) = s
\\
\ceff_s(\gamma \conc \seq{m}) = \eff(m,\ceff_s(\gamma))\;.
\end{ldispl}%
We define, for each $s \in S$, an \emph{execution architecture service}
$\funct{H_s}{\neseqof{\Meth}}{\set{\True,\False,\Div}}$
in terms of $\ceff_s$ and $\yld$ as follows:
\begin{ldispl}
H_s(\gamma \conc \seq{m}) = \yld(m,\ceff_s(\gamma))\;.
\end{ldispl}%

For each $s \in S$, $H_s$ is a service indeed:
$H_s(\gamma) = \Div \Implies H_s(\gamma \conc \seq{m}) = \Div$
for all $\gamma \in \neseqof{\Meth}$ and $m \in \Meth$.
This follows from the following property of the execution architecture
for code controlled machine structures:
\begin{ldispl}
\Exists{s \in S}
 {\Forall{i \in \Instr}{}}
\\ \quad
   (\yld(i,s) = \Div \And
     \Forall{s' \in S}
      {(\yld(i,s') = \Div \Implies \eff(i,s') = s)})\;.
\end{ldispl}%
The witnessing state of this property is $\sdiv$.
This state is connected with the undefined service $\undef$ as follows:
$H_\sdiv = \undef$.

It is worth mentioning that $H_s(\seq{m}) = \yld(m,s)$ and
$\derive{m}H_s = H_{\eff(m,s)}$.

We write $\EAS{\gM}$ for the family of services
$\set{H_s \where s \in S}$.

\section{Control Codes and Execution Architecture Services}
\label{sect-cc-exearch}

In this section, we make precise what it means that a control code is
installed on an execution architecture service and what it means that a
control code is portable from one execution architecture service to
another execution architecture service.

\subsection{Installed Control Codes}
\label{subsect-installation}

The intuition is that a control code is installed on an execution
architecture service if either some file name has assigned an executable
version of the control code or some file name has assigned an
interpretable version of the control code and an appropriate interpreter
is also installed on the execution architecture service.

Let $\gM = \tup{\Bseq,\indfam{\mf_n}{n \in \Nat},\Exec}$ be a code
controlled machine structure,
let $\tup{\CCN,\psi}$ be a control code notation for $\gM$,
let $c \in \CCN$, and let $\eas = H_{\tup{\sigma,x}} \in \EAS{\gM}$.
Then $c$ is \emph{installed} on $\eas$ if there exist
$\fnm_0,\ldots,\fnm_l \in \Fnm$ with $\sigma(\fnm_0) \in \Exec$ such
that
\begin{ldispl}
\Forall{\,\seq{\nm{bs}_1,\ldots,\nm{bs}_m} \in \seqof{\Bseq}}{{}}
\\ \;\;\;
 \AND{n \in \Nat}
  \ccmf{\gM}{n}{\psi(c)}{\nm{bs}_1,\ldots,\nm{bs}_m} =
  \ccmf{\gM}{n}
   {\sigma(\fnm_0)}
   {\sigma(\fnm_1),\ldots,\sigma(\fnm_l),\nm{bs}_1,\ldots,\nm{bs}_m}\;.
\end{ldispl}%

A control code is pre-installed on an execution architecture service if
the execution architecture service can be expanded to one on which it is
installed, using only control codes and data already assigned to file
names.
Thread algebra is brought into play to make precise what it means that
an execution architecture service can be expanded to another execution
architecture service.

Let $\gM = \tup{\Bseq,\indfam{\mf_n}{n \in \Nat},\Exec}$ be a code
controlled machine structure,
let $\eas = H_{\tup{\sigma,x}} \in \EAS{\gM}$, and
let $\eas' = H_{\tup{\sigma',x'}} \in \EAS{\gM}$.
Then $\eas$ is \emph{expansible} to $\eas'$ if:
\begin{iteml}
\item
$\dom(\sigma) \subseteq \dom(\sigma')$ and
$\sigma(\fnm) = \sigma'(\fnm)$ for all $\fnm \in \dom(\sigma)$;
\item
there exists a thread $p$ without basic actions of the form
$\ea.\seti{:}\fnm{:}\bs$ such that $\eas' = \apply{p}{\ea}{\eas}$.
\end{iteml}

Let $\gM = \tup{\Bseq,\indfam{\mf_n}{n \in \Nat},\Exec}$ be a code
controlled machine structure,
let $\tup{\CCN,\psi}$ be a control code notation for $\gM$,
let $c \in \CCN$, and let $\eas \in \EAS{\gM}$.
Then $c$ is \emph{pre-installed} on $\eas$ if
\begin{iteml}
\item
$c$ is not \emph{installed} on $\eas$;
\item
there exists a $\eas' \in \EAS{\gM}$ such that $\eas$ is expansible to
$\eas'$ and $c$ is \emph{installed} on $\eas'$.
\end{iteml}

\begin{example}
\label{example-installed-control-code}
Take an assembly code notation $\ACN$ and a source code notation $\SCN$.
Consider an execution architecture service $\eas$ on which file name
$\fnm_1$ has assigned an executable version of an assembler for $\ACN$,
file name $\fnm_2$ has assigned an $\ACN$ version of a compiler for
$\SCN$, and file name $\fnm_3$ has nothing assigned.
Suppose that no file name has assigned an executable version of the
compiler.
Then the compiler is not installed on $\eas$.
However, the compiler is pre-installed on $\eas$ because it is installed
on the expanded execution architecture service
$\apply
  {(\ea.\loai{:}\fnm_1 \bapf \ea.\exei{:}\fnm_2\iosep\fnm_3)}
  {\ea}
  {\eas}$.
\end{example}

\subsection{Portable Control Codes}
\label{subsect-portability}

We take portability of control code to mean portability from a service
defined by the execution architecture for one machine structure to a
service defined by the execution architecture for another machine
structure.

Transportability is considered a property of all bit sequences, i.e.\
each bit sequence can be transported between any two services defined by
execution architectures for machine structures.
Therefore, it is assumed that every bit sequence assigned to a file name
on a service can be assigned to a file name on another service by means
of an instruction of the form $\seti{:}\fnm{:}\bs$.

A prerequisite for portability of a control code from a service defined
by the execution architecture for one machine structure to a service
defined by the execution architecture for another machine structure is
that, for all inputs covered by the former machine structure, the
outputs produced under control of the control code coincide for the two
machine structures concerned.
Moreover, it must be possible to expand the service from which the
control code originates such that the control code is pre-installed
on the other service after some bit sequences assigned to file names on
the expanded service are assigned to file names on the other service.

Let $\gM  = \tup{\Bseq,\indfam{\mf_n}{n \in \Nat},\Exec}$ and
$\gM' = \tup{\Bseq',\indfam{\mf'_n}{n \in \Nat},\Exec'}$ be code
controlled machine structures such that $\Bseq \subseteq \Bseq'$,
let $\tup{\CCN,\psi}$ and $\tup{\CCN,\psi'}$ be control code notations
for $\gM$ and $\gM'$, respectively,
let $c \in \CCN$, and
let $\eas_0  = H_{\tup{\sigma_0,x_0}} \in \EAS{\gM}$
and $\eas'_0 = H'_{\tup{\sigma'_0,x'_0}} \in \EAS{\gM'}$.
Then $c$ is \emph{portable} from $\eas_0$ to $\eas'_0$ if
\begin{iteml}
\item
$
\begin{array}[t]{@{}l@{}}
\Forall{\,\seq{\nm{bs}_1,\ldots,\nm{bs}_m} \in \seqof{\Bseq}}{{}}
\\ \quad
 (\ccmf{\gM}{1}{\psi(c)}{\nm{bs}_1,\ldots,\nm{bs}_m} \neq \Div
\\ \quad \phantom{(} {} \Implies
  \AND{n \in \Nat}
   \ccmf{\gM}{n}{\psi(c)}{\nm{bs}_1,\ldots,\nm{bs}_m} =
   \ccmf{\gM'}{n}{\psi'(c)}{\nm{bs}_1,\ldots,\nm{bs}_m})\;.
\end{array}
$
\item
there exists a $\eas_1 = H_{\tup{\sigma_1,x_1}} \in \EAS{\gM}$ such that
\begin{iteml}
\item
$\eas_0$ is expansible to $\eas_1$,
\item
there exist
$\fnm_1,\ldots,\fnm_l \in \dom(\sigma_1) \diff \dom(\sigma'_0)$
such that $c$ is pre-installed on
$\apply
  {(\ea.\seti{:}\fnm_1{:}\sigma_1(\fnm_1) \bapf \ldots \bapf
    \ea.\seti{:}\fnm_l{:}\sigma_1(\fnm_l))}
  {\ea}
  {\eas'_0}$.
\end{iteml}
\end{iteml}

Because we assume that the set $\Fnm$ of file names is countably
infinite, this definition does not imply that the bit sequences to be
transported have to be assigned to the same file names at both sides.

\begin{example}
\label{example-portable-control-code}
Take a source code notation $\SCN$ and an assembly code notation $\ACN$.
Consider an execution architecture service $\eas$ on which file name
$\fnm_1$ has assigned an executable version of a compiler for $\SCN$
that produces assembly codes from $\ACN$, file name $\fnm_2$ has
assigned a source code from $\SCN$, and file name $\fnm_3$ has nothing
assigned.
Moreover, consider another execution architecture service $\eas'$ on
which file name $\fnm_1$ has assigned an executable version of an
assembler for $\ACN$, and file name $\fnm_3$ has nothing assigned.
Suppose that the above-mentioned prerequisite for portability of the
source code is fulfilled.
Then the source code is portable from $\eas$ to $\eas'$ because it is
pre-installed on $\apply{\ea.\seti{:}\fnm_3{:}\nm{bs}}{\ea}{\eas'}$
where $\nm{bs}$ is the bit sequence assigned to $\fnm_3$ on
$\apply
  {(\ea.\loai{:}\fnm_1 \bapf \ea.\exei{:}\fnm_2\iosep\fnm_3)}
  {\ea}
  {\eas}$.
\end{example}

\section{Conclusions}
\label{sect-concl}

We have presented a logical approach to explain issues concerning
control codes that are independent of the details of the behaviours that
are controlled at a very abstract level.
We have illustrated the approach by means of examples which demonstrate
that there are non-trivial issues that can be explained at this level.
In the explanations given, we have consciously been guided by
empirical viewpoints usually taken by practitioners rather than
theoretical viewpoints.
The issues that have been considered are well understood for quite a
time.
Application of the approach to issues that are not yet well understood
is left for future work.
We think among other things of applications in the areas of software
asset sourcing, which is an important part of IT sourcing, and software
patents.
At least the concept of control code can be exploited to put an end to
the lack of conceptual clarity in these areas about what is software.

We have based the approach on abstract machine models, called machine
structures.
If systems that provide execution environments for the executable codes
of machine structures are involved in the issues to be explained, then
more is needed.
We have introduced an execution architecture for machine structures as a
model of such systems and have explained portability of control codes
using this execution architecture and an extension of basic thread
algebra.
The execution architecture for machine structures, as well as the
extension of basic thread algebra, may form part of a setting in which
the different kinds of processes that are often transferred when
sourcing software assets, in particular software exploitation, can be
described and discussed.

We have looked at viewpoints of practitioners from a theoretical
perspective.
Unfortunately, it is unavoidable that the concepts introduced cannot all
be associated directly with the practice that we are concerned about.
This means that reading of the paper might be difficult for
practitioners.
Therefore, the paper must be considered a paper for theorists.

We have explained issues originating from the areas of compilers and
software portability.
The literature on compilers is mainly concerned with theory and
techniques of compiler construction.
A lot of that has been brought together in textbooks such
as~\cite{AU77a,Wir96a}.
To our knowledge, the phenomenon that we call the compiler fixed point
is not even informally discussed in the literature on compilers.
The literature on software portability is mainly concerned with tools,
techniques and guidelines to achieve portability.
The best-known papers on software portability are early papers such
as~\cite{PW75a,TKB78a}.
To our knowledge, the concept of portable program is only very
informally discussed in the literature on software portability.
Moreover, we are not aware of formal descriptions of compiler fixed
point and portable program in the literature on formal methods.

\acknowledgement
{This research was carried out as part of the Jacquard-project
 Symbiosis, which is funded by the Netherlands Organisation for
 Scientific Research (NWO).
}

\bibliographystyle{spmpsci}
\bibliography{CC}

\end{document}